\documentclass[journal=jctcce,manuscript=article]{achemso}
\usepackage{amsmath, amssymb, amsfonts}
\usepackage{array, booktabs, tabularx}
\usepackage{xcolor, graphicx, tikz}
\usepackage{hyperref}  
\usepackage{braket}
\usepackage{soul}
\usepackage{bm}
\usepackage{float}
\usepackage[caption = false]{subfig}
\usepackage{cancel}
\usepackage[ruled]{algorithm2e}
\usepackage[normalem]{ulem} 
\usepackage{qcircuit}
\hypersetup{colorlinks = True, allcolors = {blue}}
  
\usepackage{array}
\newcolumntype{P}[1]{>{\centering\arraybackslash}p{#1}}


\newcommand{\swap}{{\rm SWAP}}
\newcommand{\cnot}[2]{{\rm C}_{#1}{\rm NOT}_{#2}}
\newcommand{\nn}{\nonumber}

\newcommand{\YaleQI}{Yale Quantum Institute, Yale University, New Haven, CT 06511, USA}
\newcommand{\YaleChem}{Department of Chemistry, Yale University, New Haven, CT 06520, USA}
\newcommand{\LafChem}{Department of Chemistry, Lafayette College, Easton, PA 18042, USA}
\newcommand{\LafECE}{Department of Electrical and Computer Engineering, Lafayette College, Easton, PA 18042, USA}
\newcommand{\NCSUECE}{Department of Electrical and Computer Engineering, North Carolina State University, Raleigh, NC 27606, USA}
\newcommand{\NCSUCS}{Department of Computer Science, North Carolina State University, Raleigh, NC 27606, USA}
\newcommand{\NCSUPhys}{Department of Physics, North Carolina State University, Raleigh, NC 27606, USA}

\xyoption{line}
\newcommand{\ow}[1][-1]{\ar @{-} [0,#1]}
\newcommand{\oghost}[1]{*+<1em,.9em>{\hphantom{#1}} \ow}

\title{A Computational Framework for Simulations of Dissipative Non-Adiabatic Dynamics on Hybrid Oscillator-Qubit Quantum Devices}

\author{Nam P. Vu}
 \affiliation{\LafChem}
 \alsoaffiliation{\YaleChem}
 \alsoaffiliation{\LafECE}

\author{Daniel Dong}
 \affiliation{\NCSUCS}

\author{Xiaohan Dan}
 \affiliation{\YaleChem}
 
\author{Ningyi Lyu}
 \affiliation{\YaleChem}

\author{Victor Batista}
 \email{victor.batista@yale.edu}
 \affiliation{\YaleChem}
 \alsoaffiliation{\YaleQI}

\author{Yuan Liu}
 \email{q\_yuanliu@ncsu.edu}
 \affiliation{\NCSUECE}
 \alsoaffiliation{\NCSUCS}
 \alsoaffiliation{\NCSUPhys}

\begin{document}

\begin{abstract}
    We introduce a computational framework for simulating non-adiabatic vibronic dynamics on circuit quantum electrodynamics (cQED) platforms. Our approach leverages hybrid oscillator-qubit quantum hardware with mid-circuit measurements and resets, enabling the incorporation of environmental effects such as dissipation and dephasing. To demonstrate its capabilities, we simulate energy transfer dynamics in a triad model of photosynthetic chromophores inspired by natural antenna systems. We specifically investigate the role of dissipation during the relaxation dynamics following photoexcitation, where electronic transitions are coupled to the evolution of quantum vibrational modes. Our results indicate that hybrid oscillator-qubit devices, operating with noise levels below the intrinsic dissipation rates of typical molecular antenna systems, can achieve the simulation fidelity required for practical computations on near-term and early fault-tolerant quantum computing platforms.
\end{abstract}
\maketitle

\section{Introduction}
The complex interplay between the electronic, vibrational, and environmental degrees of freedom in organic molecules underpins efficient photosynthetic processes~\cite{arsenault2021vibronic, cao2020quantum} as well as many other charge and energy transfer phenomena, including intramolecular energy redistribution~\cite{freed1980intramolecular} and vibrational-selective chemical reactions.~\cite{pascual2003selectivity} Given the ubiquitous role of vibronic dynamics, the development of computational frameworks for efficient and accurate simulations of vibronic systems is a subject of great interest.~\cite{lim2015vibronic} Here, we introduce a computational framework for hybrid oscillator-qubit quantum hardware.

Simulating vibronic dynamics on classical computers is challenging due to the exponential growth of the Hilbert space dimension with the number of vibrational modes. Despite this, numerically exact methods have been developed to propagate quantum dynamics within a truncated Hilbert space.~\cite{zhang1998theory,tanimura2020numerically,greene2017tensor,lyu2022tensor,baiardi2019large,ren2018time,li2020numerical,shi18,yan21a,ke22,dan23b,guan24,bai24} For quantum systems with limited entanglement, state-of-the-art algorithms rely on tensor factorization methods based on matrix product state or tensor-train representations.~\cite{greene2017tensor,lyu2022tensor,baiardi2019large,ren2018time,li2020numerical,shi18,yan21a,ke22,dan23b,guan24,bai24}  These approaches enable efficient and accurate simulations by truncating the bond dimension (or Schmidt rank) to manage computational costs. Other exact methods, such as the hierarchical equations of motion (HEOM) and the pseudomode framework,~\cite{garraway97,lambert19,dan23,dan25,menczel24,bai24} simplify the problem by mapping many vibrational modes onto a smaller set of pseudomodes. This significantly reduces the Hilbert space dimension. However, these techniques are generally restricted to systems with linear couplings between electronic and vibrational degrees of freedom.

Approximate methods have also been proposed to address the computational challenges of simulating vibronic dynamics. These include mapping electronic and vibrational states to simplified representations~\cite{Bonella2001MapHam,Miller2001Map} and employing many (quasi)-classical trajectories to model dynamics at reduced computational costs.~\cite{Hornicek2007ManyClassical,Provazz2021QuClassical} However, assessing the accuracy of these methods can be challenging.~\cite{liu2024benchmarking} A recent study indicates that the choice of an optimal approximation method is highly system-dependent: simulation accuracy is influenced by several factors, including the initial sampling strategy for mapping variables.~\cite{liu2024benchmarking} This underscores the need for developing computational frameworks for efficient yet rigorous simulations.

Over the past decade, significant advances have been achieved in the engineering and control of continuous-variable (CV) bosonic quantum devices,~\cite{Braunstein2005,Weedbrook2012,andersen2015hybrid,YYGao_QIP_bosonic_cQED,liu2024hybrid} in addition to their discrete-variable (DV) counterparts.~\cite{kubica2022erasure,ReinholdErrorCorrectedGates} These breakthroughs suggest that the challenges of simulating complex polyatomic vibronic dynamics on classical computers could be addressed by mapping molecular vibrations onto native bosonic hardware.~\cite{WangConicalIntersection,Whitlow2023,dutta2024chemistry} With universal control on hybrid oscillator-qubit platforms,~\cite{QuditsfromOscillatorsPhysRevA.104.032605,EickbuschECD,liu2024hybrid,liu2024hybrid} the quantum dynamics of any vibronic Hamiltonian can be simulated, in principle, given sufficiently many high-fidelity bosonic modes.

However, several challenges must be addressed to effectively utilize hybrid oscillator-qubit quantum hardware for realistic vibronic simulations. First, the limited connectivity and native gate sets on current quantum hardware raise questions about the computational overhead required to map and compile the Hamiltonian for near-term devices. Second, realistic vibronic dynamics are inherently non-unitary~\cite{bao2024vibronic} due to dissipation induced by the surrounding environment. This calls for the development of systematic approaches to simulate dissipative quantum dynamics on hybrid CV-DV platforms. Finally, quantum hardware is inherently susceptible to noise.~\cite{Georgopoulos2021Noise,DiBartolomeo2023noise} The impacts of intrinsic noise on the accuracy and feasibility of quantum simulations using near-term hybrid CV-DV devices remain unclear and require further investigation. 

In the present work, we address these challenges by co-designing scalable, near-term hybrid oscillator-qubit quantum modular hardware for simulating dissipative vibronic dynamics, the first of its kind to the best of our knowledge. We focus on the bosonic circuit quantum electrodynamics (cQED) platform \cite{BlaiscQEDReviewRMP2020,YYGao_QIP_bosonic_cQED} as a case study, yet the approach is broadly applicable to other quantum hardware platforms equipped with native bosonic modes and qubits. We provide a concrete mapping and quantum circuit realization of the dissipative dynamics using a native instruction set architecture for cQED hardware. Additionally, we present a unitary method to simulate Markovian dephasing and amplitude damping processes by appropriately engineering quantum channels for cQED hardware modules. A detailed gate count for resource estimation is also included, along with an analysis on how various intrinsic cQED hardware noise impact the simulation results. To validate our approach, we perform numerical simulations of energy transfer dynamics in a three-site chromophore antenna model. The results highlight the importance of dissipation in energy transfer dynamics, where we demonstrate how tuning amplitude damping rates on specific chromophores can significantly alter the dominant energy transfer pathway.

The structure of the paper is organized as follows. Sec.~\ref{sec:methods} introduces the Hamiltonian and dissipation model for a photosynthetic antenna model composed of a one-dimensional chromophore array. Sec.~\ref{sec:results} presents the main findings, focusing on the co-design of quantum circuit and layouts to simulate chromophore dynamics using native operations on cQED hardware. Sec.~\ref{sec:results} provides extensive numerical simulations to validate the proposed circuits and explore the role of dissipation in energy transfer dynamics. Sec.~\ref{sec:conclusion} concludes the paper with future outlooks and potential research directions.

\section{Methods}\label{sec:methods}
This section is organized as follows. Sec.~\ref{ssec:vibronic-Hamiltonian} introduces the vibronic Hamiltonian for a model photosynthetic antenna and showcases its cQED formulation. Sec.~\ref{ssec:energy-transfer-problem} discusses the energy transfer problem of interest. Sec.~\ref{ssec:engineering-dissipation} describes our approach for engineering environment-induced dissipation via channel dilation techniques. We then propose our cQED modular hardware design in Sec.~\ref{ssec:exp}, followed by quantum circuit realization with resource estimation to simulate vibronic dynamics in Sec.~\ref{ssec:ham-sim}.

\subsection{Photosynthetic Model}\label{ssec:vibronic-Hamiltonian}

\subsubsection{Vibronic Hamiltonian Model}
\begin{figure}[htbp]
    \centering
    \subfloat[Photosynthesis Model]{\includegraphics[width=0.48\textwidth]{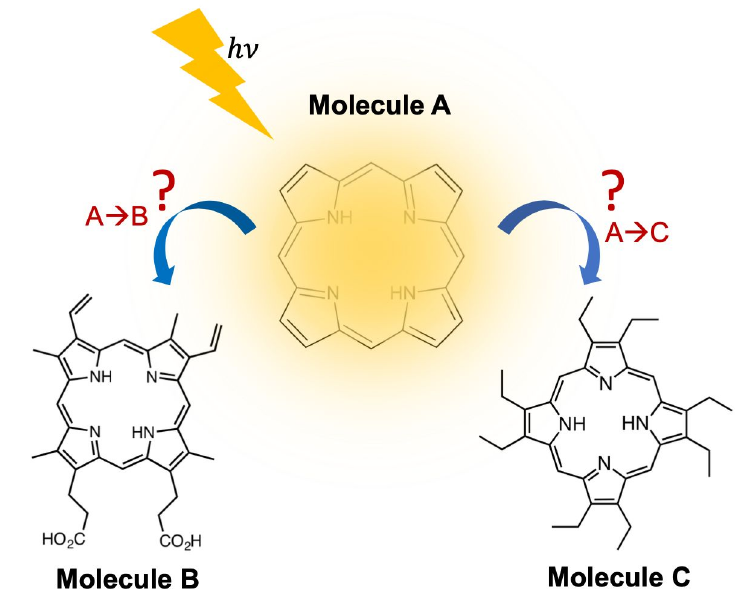}}
    \subfloat[Proposed Hardware Layout]{\includegraphics[width=0.48\textwidth]{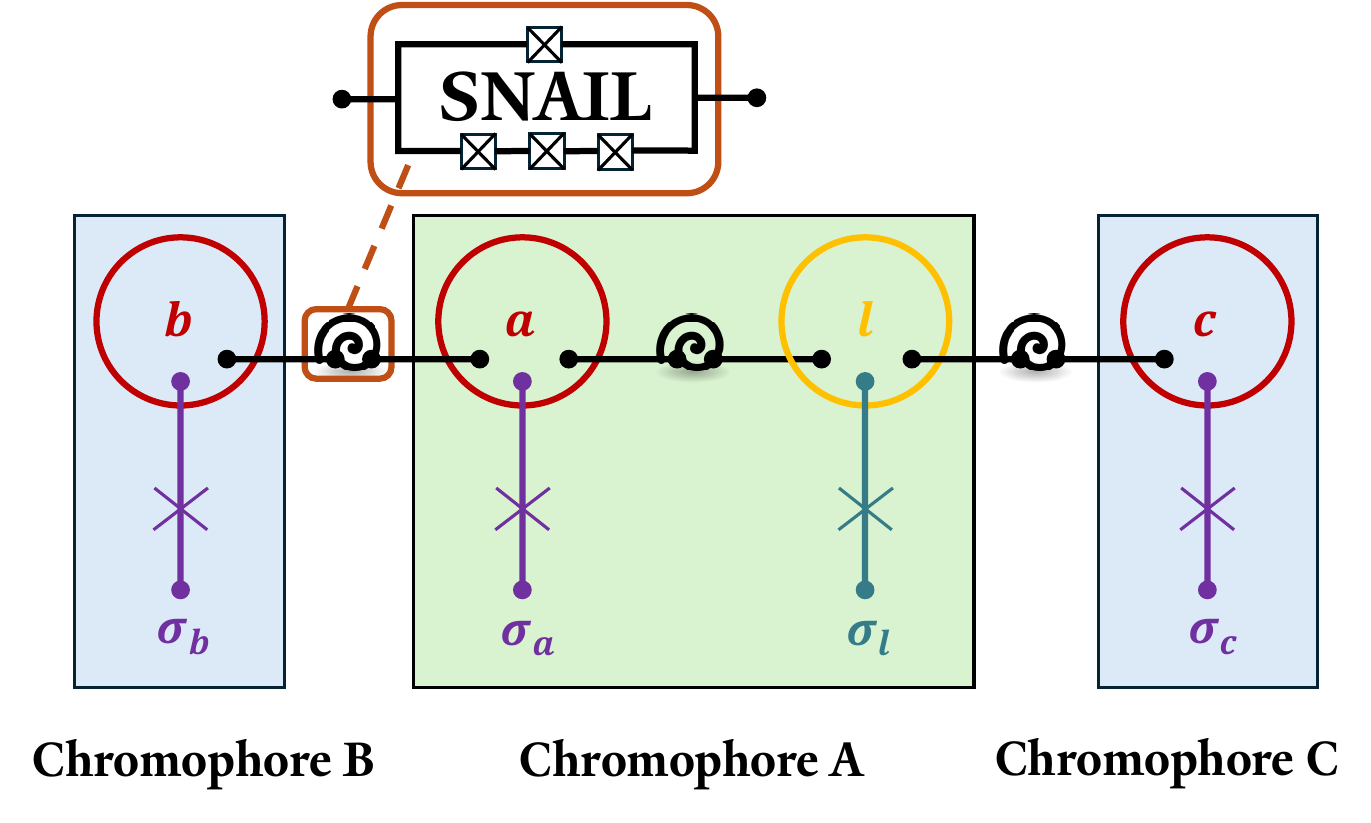}}
    \caption{(a) Photosynthetic antenna model system, composed of three chromophores representing distinct pigments within a protein. One elementary problem is to determine if an initial electronic excitation on chromophore $A$ has a dominant energy transfer pathway, and if so, whether it favors energy transfer to chromophore $B$ or $C$. (b) Proposed cQED modular hardware for simulating vibronic dynamics of a three-site chromophore system. High-frequency (red circles) and low-frequency (yellow circles) cavities represent vibrational modes. A SNAIL device mediates coupling between adjacent cavities. High-frequency cavities are coupled to transmon qubits (shown in purple), representing the ground and excited electronic states of each chromophore, while ancillary qubits for low-frequency cavities are shown in teal blue.}
    \label{fig:3site-proposal}
\end{figure}

We consider the model system illustrated in Fig.~\ref{fig:3site-proposal} (a) which consists of three chromophores labeled as sites $A, B,$ and $C$. In the context of photosynthetic antennas, these chromophores represent distinct pigments within a protein, as modeled in Ref.~\citenum{arsenault2021vibronic}. Each chromophore has one electronic degree-of-freedom (i.e., a two-level system representing ground and excited electronic states) coupled to one high-frequency vibrational mode (labeled as $a,b,c$) and only interacts with its adjacent chromophores. These high-frequency modes represent local vibrations, such as bond stretching or bending, of which the frequencies and equilibrium positions are specific to each chromophore. 

Additionally, chromophore $A$ also has a low-frequency vibrational mode $l$, whose equilibrium position depends on the state of chromophore $A$. Mode $l$ can intuitively be interpreted as a long-wavelength, vibrational coordinate that strongly couples to two or more chromophores. Furthermore, the (electronically) excited state of chromophore $A$ is dipole-coupled to the excited states of chromophores $B$ and $C$. These couplings, with strengths $J_{AB}$ and $J_{AC}$, are modulated differently by the coordinates of mode $l$. 

The photochemistry process is shown in Fig.~\ref{fig:3site-proposal} (a). Initially, chromophore $A$'s electronic state is excited. The excitation energy is then transferred to chromophores $B$ or $C$ at rates described by the coupling constants $J_{AB}$ and $J_{AC}$. The excited chromophores $B$ or $C$ can also transfer energy back to $A$ at the same rates. Local vibronic coupling in each chromophore facilitates electronic-to-vibrational energy transfer.

We restrict the system Hamiltonian to the ground state and singly excited state manifold. In this construction, at most one of the three chromophores can be excited at a time, while the others remain in the ground state. Thus, double-excitations and triple-excitations are excluded by design. We denote the ground and excited states of an individual chromophore as $s = g, e$, respectively. The state $\ket{G} = \ket{g_A g_B g_C}$ represents all chromophores in their ground electronic states. The state $\ket{R}$ indicates that the chromophore $R = A, B, C$ is in its excited state while the others are in their ground states. In other words, a local excitation on chromophore $A$ is written as $\ket{A} = \ket{e_A g_B g_C}$, while for chromophores $B$ and $C$, the respective excited states are $\ket{B} = \ket{g_A e_B g_C}$, and $\ket{C} = \ket{g_A g_B e_C}$.

The vibrational Hamiltonian for a chromophore $R$ in state $s = g, e$ is denoted as $h_R^{s}$. Using this notation, the full system Hamiltonian for the four possible electronic states (ground state and three singly excited states), coupled to four vibrational modes with distinct frequencies, is given by:
\begin{align}
    H=& \ket{G}\bra{G} \otimes (\hat{h}_A^g + \hat{h}_B^g + \hat{h}_C^g) + \ket{A}\bra{A} \otimes (\hat{h}_A^{e} + \hat{h}_B^{g} + \hat{h}_C^{g})  \nonumber \\
      &+ \ket{B}\bra{B} \otimes (\hat{h}_A^{g} + \hat{h}_B^{e} + \hat{h}_C^{g}) + \ket{C}\bra{C} \otimes (\hat{h}_A^{g} + \hat{h}_B^{g} + \hat{h}_C^{e})  \nonumber \\
      &+ J_{AB} \left( \ket{A}\bra{B} + h.c. \right) + J_{AC} \left( \ket{A}\bra{C} + h.c. \right). \label{general-ham-5}
      \end{align}
Here, terms 1-4 describe the Hamiltonians for configurations with at most one excited chromophore. The fifth term accounts for the dipole couplings between the excited state of chromophore $A$ and those of $B$ or $C$, with coupling constants $J_{AB}$ and $J_{AC}$, respectively.

Using the bosonic annihilation operators $a,b,c$ for the high-frequency vibrational modes of chromophores $A, B$ and $C$, and $l$ for the low-frequency vibrational mode of chromophore $A$, the vibronic Hamiltonians are defined, as follows. For chromophore $A$,
\begin{align}
    \hat{h}^g_A &= \hbar \omega_{g, a} (a^\dagger a + \frac{1}{2}) + \hbar \omega_{l} (l^\dagger l + \frac{1}{2}), \label{eq:general_hgA} \\
    \hat{h}^e_A &= \hbar \omega_{e, a} \left[ a^\dagger a + \frac{1}{2} + S_a - \sqrt{S_a} (a^\dagger + a) \right] + \hbar \omega_{l} \left[ l^\dagger l + \frac{1}{2} + S_l - \sqrt{S_l} (l^\dagger + l) \right]. \label{eq:general_heA}
\end{align}
For chromophores $B$ and $C$ (with $R = B, C$ and $r = b, c$), 
\begin{align}
    \hat{h}^g_R &= \hbar \omega_{g, r} (r^\dagger r + \frac{1}{2}), \label{eq:general_hgR} \\
    \hat{h}^e_R &= \hbar \omega_{e, r} \left[r^\dagger r + \frac{1}{2} + S_r - \sqrt{S_r} (r^\dagger + r) \right]. \label{eq:general_heR}
\end{align}
Here, $S_r$ represents the Huang-Rhys factors, which characterize the vibronic coupling strengths.~\cite{Huang1950HRFactor} The dipole coupling constants between chromophores $A$ and $R$ ($R =  B, C$) are given by:
\begin{align}
    J_{AR} = \hbar J_{AR,0} \left[1 + \eta_{AR} (l^\dagger + l) \right], \label{eq:general_ham_JAR}
\end{align}
where $\eta_{AR}$ is a first-order coupling constant. The modulation of $J_{AR}$ by the low-frequency mode position $(l+l^\dagger)$ reflects the natural influence of vibronic coupling. 

Additionally, the surrounding environment induces energy dissipation, including amplitude damping and dephasing effects at respective rates $\gamma_{\rm amp,all}$ and $\gamma_{\rm dep,all}$. These processes are described by the Lindblad quantum master equation as outlined in Sec.~\ref{sec:diss}. Our photosynthetic antenna model system is parametrized with physically relevant values given in Table \ref{tab:params}.


\subsubsection{Dissipative Dynamics} \label{sec:diss}
In this subsection, we describe how energy dissipation from the excited chromophore population, under the influence of environmental effects, can be modeled using the spin-boson model and Lindblad master equation.

Each chromophore's electronic state is modeled as a two-level quantum system described by $H_S$ and interacts with its surrounding environment according to the following spin-boson Hamiltonian:
\begin{equation} \label{eq:HT}
    {H}_T = H_S + 
\sum_a \frac{1}{2}\left[p_a^2 + \omega_a^2\left(x_a - \frac{c_a}{\omega_a^2}O_S\right)^2\right].
\end{equation}
Here, $H_S = -E_0 \sigma^z$ is the system Hamiltonian, with $E_0$ is the energy difference between the excited state $\ket{e}$ and ground state $\ket{g}$, while $H_B = \frac{1}{2} \sum_a (p_a^2 + \omega_a^2 x_a^2)$ is the harmonic bath Hamiltonian. On the other hand, $H_I = -O_S \sum_a c_a x_a$ is the coupling between system and bath, with the arbitrary operator in the system Hilbert space $O_S = \eta_x \sigma^x + \eta_y \sigma^y + \eta_z \sigma^z + \eta_I I$ expressed as a linear combination of the four Pauli matrices.

The environmental effects are captured by the coupling constants $c_a$, introduced in Eq.~\eqref{eq:HT}, that can be obtained from the reservoir correlation function:~\cite{breuer02, ishizaki09a, weiss12}
\begin{equation}
    C(t) = \frac{1}{\pi} \int_0^\infty d\omega \: J(\omega) \left[\coth\left(\frac{\beta \omega}{2}\right) \cos(\omega t) - i \sin(\omega t)\right],
\end{equation}
where $J(\omega) = \frac{\pi}{2}\sum_a \frac{c_a^2}{\omega_a} \delta(\omega - \omega_a)$ is the bath spectral density and $\beta = 1/kT$ is the inverse temperature.
 
Assuming the Born-Markov and rotating wave approximations (RWA), the dynamics can be described by the evolution of the reduced density matrix of the system, $\rho(t) = \mathrm{Tr}_B[\rho_T(t)]$, according to the Lindblad equation:~\cite{manzano20} \begin{equation}
    \frac{d\rho(t)}{dt} = -i [H_S, \rho(t)] + 
     \sum_\omega \gamma(\omega) 
    \Big( L(\omega)\rho(t) L^\dagger(\omega) - \frac{1}{2}\{L^\dagger(\omega)L(\omega), \rho(t)\} \Big),
\end{equation}
where $\gamma(\omega) = \frac{2J(\omega)}{1 - e^{-\beta \omega}}$ is the damping rate, and \begin{equation}
    L(\omega) = \sum_{\epsilon' - \epsilon = \omega} \langle \epsilon |O_S|\epsilon'\rangle |\epsilon\rangle \langle \epsilon'|
\end{equation}
are the jump operators in the eigenbasis of $H_S$, comprised of $\ket{0}$ and $\ket{1}$ with respective eigenvalues $-E_0$ and $E_0$, i.e., $H_S \ket{0} = -E_0 \ket{0}$ and $H_S \ket{1} = E_0 \ket{1}$.
Substituting these eigenstates into $L(\omega)$, we obtain three primary jump operators, corresponding to: \begin{equation}
    L(\omega = 0) = \eta_z\sigma^z, \quad\quad L(\omega = \pm 2E_0) = \left(\eta_x \mp i\eta_y \right) \sigma^\pm.
\end{equation}
Here, $\sigma^\pm = (\sigma^x \pm i\sigma^y)/2$ are the raising and lowering operators and thus respectively represent the relaxation and excitation environmental effects, while the $\sigma^z$ operator describes pure dephasing. Table~\ref{tab:para_lindblad_SBM} summarizes the derived jump operators alongside their damping rates. 
\begin{table}[H]
    \centering
    \begin{tabular}{|c|c|c|}
    \hline
    \textbf{Process} & \textbf{Jump operator} & \textbf{Dissipation rate} \\ \hline
    \text{Relaxation} & $\sigma^+$ & $2(\eta_x^2 + \eta_y^2)\frac{J(2 E_0)}{1-e^{-2\beta E_0}}$  \\
    \text{Excitation} & $\sigma^-$ & $2(\eta_x^2 + \eta_y^2)\frac{J(-2 E_0)}{1-e^{2\beta E_0}}$ \\
    \text{Dephasing} & $\sigma^z$ & $\eta_z^2\frac{J'(0)}{\beta}$  \\
    \hline
\end{tabular}
    \caption{Parameters for the Lindblad equation as derived for the spin-boson model. The term $J'(0) =$ $\frac{\partial}{\partial \omega} J(\omega) \big|_{\omega=0}$ represents the first derivative of the spectral density $J(\omega)$ at $\omega = 0$. We note that energy absorption from the environment, described by the jump operator $\sigma^-$, becomes abysmal when $E_0 \gg kT$.}
    \label{tab:para_lindblad_SBM}
\end{table}
By accurately characterizing environmental effects, this dynamical model provides a comprehensive description of essential quantum energy transfer processes in our photosynthetic antenna model. We also note that the Lindblad equation, as implemented in this study, has been widely used to describe dissipation in a wide range of contexts, including quantum information science.~\cite{chen2022hamiltonian} However, it is based on several approximations that limit its applicability to systems that are weakly coupled to their environment.\cite{breuer02,mohseni08,xu17,xu18a,manzano20,dan22,dan25} Hence, more rigorous quantum master equations should be used when its applicability is exceeded.

\subsubsection{Effective Hamiltonian of the cQED Platform} \label{sec:ehcqed}
The cQED platform, shown in Fig.~\ref{fig:3site-proposal} (b), enables the simulation of the model system shown in Fig.~\ref{fig:3site-proposal} (a), upon suitable parametrization of the quantum operations applied. Each microwave cavity of the device corresponds to a vibrational mode of the chromophores, while the ground and excited electronic states of chromophores $A$, $B$, and $C$ are mapped onto the ground $\ket{0}$ and excited $\ket{1}$ states of qubits $\sigma_a$, $\sigma_b$ and $\sigma_c$, respectively. Therefore, within the single-excitation manifold, {\allowdisplaybreaks\begin{align}
    \ket{G} &\mapsto \ket{0}_a\otimes \ket{0}_b\otimes \ket{0}_c, \quad \quad \ket{A} \mapsto \ket{1}_a \otimes \ket{0}_b \otimes \ket{0}_c, \nn \\
    \ket{B} &\mapsto \ket{0}_a\otimes\ket{1}_b\otimes\ket{0}_c, \quad\quad \ket{C} \mapsto \ket{0}_a\otimes \ket{0}_b\otimes\ket{1}_c.
\end{align}}

Appendix~\ref{asec:H-derivation} shows how the system Hamiltonian $H$ in Eq.~\ref{general-ham-5} can be unitarily transformed into the following effective Hamiltonian in the rotating frame:
\begin{equation} \label{eq:final-ham-3site}
    \tilde{H}/\hbar = \tilde{H}_0/\hbar + \tilde{H}_1/\hbar + \tilde{H}_{2,XX}/\hbar + \tilde{H}_{2, YY}/\hbar.
\end{equation}
The four terms of $\tilde{H}$ are defined as follows:
\begin{enumerate}
    \item {\em Base Hamiltonian} ($\tilde{H}_0$): \begin{equation}
        \tilde{H}_0/\hbar 
        = \omega_a a^\dagger a + \omega_b b^\dagger b + \omega_c c^\dagger c + \omega_l l^\dagger l - \Delta_{ab} \frac{\sigma_b^z}{2} - \Delta_{ac} \frac{\sigma_c^z}{2}, \label{eq:final-ham-3site-h0}
    \end{equation}
    \item {\em Interaction Terms} ($\tilde{H}_1$): \begin{align}
        \tilde{H}_1/\hbar
        &= -\frac{\chi_a}{2} a^\dagger a \sigma_a^z - \frac{\chi_b}{2} b^\dagger b \sigma_b^z - \frac{\chi_c}{2} c^\dagger c \sigma_c^z \nn \\ 
        &+ g_{cd,a} (a + a^\dagger)\frac{\sigma_a^z}{2} + g_{cd,b} (b + b^\dagger)\frac{\sigma_b^z}{2} + g_{cd,c} (c + c^\dagger)\frac{\sigma_c^z}{2}, \label{eq:final-ham-3site-h1} 
    \end{align}
    \item {\em XX Coupling} ($\tilde{H}_{2,XX}$):\begin{align}     
        \tilde{H}_{2,XX}/\hbar
        &=  g_{cd,l} (l + l^\dagger)\frac{\sigma_a^z}{4} + \frac{g_{ab}}{2} (\sigma_a^x \sigma_b^x) + \frac{g_{ac}}{2} (\sigma_a^x \sigma_c^x) \nonumber
        \\&+ \frac{g_{abl}}{2} (\sigma_a^x \sigma_b^x )(l + l^\dagger) + \frac{g_{acl}}{2} (\sigma_a^x \sigma_c^x ) (l + l^\dagger),
         \label{eq:final-ham-3site-h2xx}
    \end{align}
    \item {\em YY Coupling} ($\tilde{H}_{2,YY}$):{\allowdisplaybreaks \begin{align}
        \tilde{H}_{2,YY}/\hbar &
        =  g_{cd,l} (l + l^\dagger)\frac{\sigma_a^z}{4} + \frac{g_{ab}}{2} (\sigma_a^y \sigma_b^y) + \frac{g_{ac}}{2} ( \sigma_a^y \sigma_c^y) \nonumber \\
        &+ \frac{g_{abl}}{2} (\sigma_a^y \sigma_b^y )(l + l^\dagger) + \frac{g_{acl}}{2} (\sigma_a^y \sigma_c^y) (l + l^\dagger). \label{eq:final-ham-3site-h2yy}
\end{align}}
\end{enumerate}

\subsubsection{Hamiltonian for a 1D Array of Coupled Chromophores}
In a more realistic photosynthetic setting, we consider multiple chromophores coupled together in a one-dimensional (1D) array, where every three neighboring chromophores interact according to the Hamiltonian described previously. Let each chromophore be labeled by the index $\xi$, each having both a high-frequency mode $\xi_0$ and a low-frequency vibrational mode $\xi_1$. Using $b_{\xi_{0(1)}}$ and $b_{\xi_{0(1)}}^\dag$ to represent the bosonic creation and annihilation operators for the high(low)-frequency mode of the $\xi^\text{th}$ chromophore, the overall Hamiltonian of an $N$-chromophore 1D array of chromophores can be written in the form 
\begin{equation} \label{total-ham-1d}
    H = \sum_{\xi = 1}^N H_0^{(\xi)} + H_1^{(\xi)} + H_2^{(\xi)}, 
\end{equation}
where the non-interacting part of the Hamiltonian, 
\begin{equation}
    H_0^{(\xi)} =  \quad \omega_{\xi_0} b^\dagger_{\xi_0} b_{\xi_0} + \omega_{\xi_1} b^\dagger_{\xi_1} b_{\xi_1} -\frac{\omega_{q\xi_0}}{2} \sigma^z_{\xi_0}, \label{eq:final-ham-msite-h0}
\end{equation}
describes the free evolution of the vibrational modes the electronic states of each chromophore. The dispersive interactions within each chromophore, primarily involving the high-frequency mode, are captured by, \begin{equation}
    H_1^{(\xi)} = -\frac{\chi_{\xi_0}}{2} b^\dagger_{\xi_0} b_{\xi_0} \sigma^z_{\xi_0} + \frac{g_{cd, \xi_0}}{2} (b_{\xi_0} + b^\dagger_{\xi_0}) \sigma^z_{\xi_0} \label{eq:final-ham-msite-h1}.
\end{equation} 
Finally, the inter-chromophore and intra-chromophore couplings, involving interactions between vibrational modes and electronic transitions are described by
{\allowdisplaybreaks
\begin{align}
    H_2^{(\xi)} &= \frac{g_{cd, \xi_1}}{2} (b_{\xi_1} + b^\dagger_{\xi_1})\sigma^z_{\xi_0} + \frac{g_{\xi_0, (\xi-1)_0}}{2} (\sigma^+_{\xi_0} \sigma^-_{(\xi-1)_0}  +h.c.) + \frac{g_{\xi_0, (\xi+1)_0}}{2} (\sigma^+_{\xi_0} \sigma^-_{(\xi+1)_0} + h.c.)\nn \\
                &+ \frac{g_{\xi_0, (\xi-1)_0, \xi_1}}{2} (\sigma^+_{\xi_0} \sigma^-_{(\xi-1)_0} +h.c.) (b_{\xi_1} + b^\dagger_{\xi_1}) + \frac{g_{\xi_0, (\xi+1)_0, \xi_1}}{2} (\sigma^+_{\xi_0} \sigma^-_{(\xi+1)_0} +h.c.)(b_{\xi_1} + b^\dagger_{\xi_1}),
                    \label{eq:final-ham-msite-h2}
\end{align}}
where the first line describes the coupling between the low-frequency vibrational mode and the electronic state within the same chromophore. The second and third lines represent {\em nearest-neighbor electronic couplings} between adjacent chromophores. The fourth and fifth lines describe {\em vibronic couplings}, where inter-chromophore electronic transitions are modulated by the low-frequency vibrational modes.

It is important to note that all inter-chromophore interaction coefficients in Eq.~\eqref{eq:final-ham-msite-h2} are divided by a factor of 2 compared to Eqs.~\eqref{eq:final-ham-3site-h2xx} and \eqref{eq:final-ham-3site-h2yy} to avoid double counting interactions. Additionally, each qubit drive frequency $\omega_{q\xi_0}$ can be obtained, following a relationship analogous to how $\omega_{qa}$ is obtained from $\omega_l$ in Eq.~\ref{eq:asec_Ha_final} and Table~\ref{tab:params-final-H}. This dependency reflects the influence of the vibrational modes on the chromophore electronic states.

\subsection{Energy Transfer Mechanism} \label{ssec:energy-transfer-problem}
Understanding the dominant energy transfer pathways in photosynthetic systems is a fundamental problem with significant implications for both natural and artificial light-harvesting processes. An illustrative example for a three-chromophore system is shown in Fig.~\ref{fig:3site-proposal} (a).~\cite{schlau2010spectroscopic,fleming2023photosynthetic} Consider an electronic excitation initially generated on molecule $A$ through photoexcitation by sunlight. As molecule $A$ is coupled to both molecules $B$ and $C$, excitation energy can transfer between these adjacent sites. The rates of energy transfer rates are determined by the specific chemical interactions as described by the corresponding coupling coefficients. The absorbed sunlight energy is subsequently used to drive downstream chemical reactions associated with charge separation and water oxidation.~\cite{shen2015structure} Thereore, it is of great interest to understand how energy transfers through specific relaxation pathways that are determined by chemical interactions, quantum interference, and dissipation.

Traditionally, tackling this problem requires solving the quantum master equation (QME),~\cite{Kelly2016GQME, brian2021generalized, Liu2024GQME} which poses significant computational challenges, especially for complex systems with a large number of degrees of freedom.~\cite{Campaioli2024QME} The situation becomes even more demanding when the {\em vibrational modes} must be treated quantum mechanically, as accurate simulation of such {\em bosonic quantum dynamics} is computationally intensive on both classical and qubit-based quantum hardware.~\cite{crane2024hybrid,wang2023simulating} 
Given the rapid advancements in cQED hardware, we propose an alternative approach that leverages the mapping of system Hamiltonians onto cQED hardware modules, integrated with novel quantum algorithms and advanced simulation techniques. This CV-DV hybrid framework offers the potential to efficiently tackle the energy transfer pathway problem, providing deeper insights into the fundamental mechanisms governing photosynthetic energy conversion.

\subsection{Engineering Dissipation Channels for Chromophores} \label{ssec:engineering-dissipation}
The dissipation dynamics described by the Lindblad equation in Sec.~\ref{sec:diss} can be modeled using damping and dephasing channels, represented by the jump operators $\sigma^+$, $\sigma^-$, and $\sigma^z$. In this section, we show how the damping rates listed in Table~\ref{tab:para_lindblad_SBM} correspond to the quantum circuit parameters for the Markovian dissipative channels derived in Appendix~\ref{asec:channel-derivation}. This connection enables the simulation of colored bath effects on the system qubit via ancilla qubits, employing unitary dilation techniques.

\begin{figure}[H]
    \centering
    \subfloat[Excitation Channel]{
        \mbox{
            \Qcircuit @C=1em @R=0.1em {
                \lstick{\ket{\phi}} & \multigate{2}{\mathcal{E}_{\rm exc}} & \qw &&&\gate{X} & \ctrl{2} &\gate{X}& \targ & \qw & \qw \\
                &\nghost{\mathcal{E}_{\rm exc}}&&=&&&&&&&\\
                \lstick{\ket{0}} & \ghost{\mathcal{E}_{\rm exc}} & \qw &&& \qw & \gate{R_y(\theta)} &\qw & \ctrl{-2} & \meter & \qw
            }
        }
    }
    \\
    \subfloat[Amplitude Damping Channel]{
        \mbox{
            \Qcircuit @C=0.9em @R=0.1em {
                \lstick{\ket{\phi}} & \multigate{2}{\mathcal{E}_{\rm amp}} & \qw &&& \ctrl{2} & \targ & \qw & \qw \\
                &\nghost{\mathcal{E}_{\rm amp}}&&=&&&&&&&&\\
                \lstick{\ket{0}} & \ghost{\mathcal{E}_{\rm amp}} & \qw &&& \gate{R_y(\theta)}& \ctrl{-2} & \meter & \qw
            }
        }
    }
    \subfloat[Dephasing Channel]{
        \mbox{
            \Qcircuit @C=0.9em @R=0.1em {
                \lstick{\ket{\phi}} & \multigate{2}{\mathcal{E}_{\rm dep}} & \qw &&& \qw & \ctrl{2} & \qw & \qw\\
                &\nghost{\mathcal{E}_{\rm dep}}&&=&&&&&&&&\\
                \lstick{\ket{0}} & \ghost{\mathcal{E}_{\rm dep}} & \qw &&& \gate{R_y(\theta)} & \gate{Z}& \meter & \qw
            }
        }
    }
    \\
    \subfloat[Spin-boson Model General Dissipation Channel]{
        \mbox{
            \Qcircuit @C=.9em @R=0.1em {
            \\ \\ \\ \\ \\ \\ \\ \\ \\ \\ \\ \\ \\
                \lstick{\ket{\phi}} & \multigate{2}{\mathcal{E}_\tau} & \qw &&& \multigate{2}{\mathcal{E}_{\rm amp}} & \multigate{2}{\mathcal{E}_{\rm exc}} & \multigate{2}{\mathcal{E}_{\rm dep}} &\qw & \\
                &\nghost{\mathcal{E}_\tau} && = && \nghost{\mathcal{E}_{\rm amp}} & \nghost{\mathcal{E}_{\rm exc}} & \nghost{\mathcal{E}_{\rm dep}} & &\\
                \lstick{\ket{0}} & \ghost{\mathcal{E}_\tau}  & \qw &&& \ghost{\mathcal{E}_{\rm amp}} & \ghost{\mathcal{E}_{\rm exc}} & \ghost{\mathcal{E}_{\rm dep}} & \qw &
            }
        }
    }
    \quad
    \subfloat[Trotter Evolution]{
        \mbox{
            \Qcircuit @C=0.9em @R=1em {
                && \text{\small $1^{st}$ Trotter layer} &&&& \text{\small $2^{nd}$ Trotter layer} &&&& \cdots \\
                \lstick{\ket{\phi}} & \gate{U_\tau} & \qw &     
                \multigate{1}{\mathcal{E}_\tau} & \qw & \gate{U_\tau} & \qw &
                \multigate{1}{\mathcal{E}_\tau} &\qw &\qw & \cdots \\
                &&  \lstick{\ket{0}} &
                \ghost{\mathcal{E}_\tau}  &\qw && \lstick{\ket{0}} &
                \ghost{\mathcal{E}_\tau} &\qw &\qw & \cdots
                \gategroup{2}{2}{3}{5}{.5em}{^\}}
                \gategroup{2}{6}{3}{9}{.5em}{^\}}
            }
        }
    }
    \quad\quad
    \caption{Quantum circuit realizations of different dissipation channels, where the system qubit $\ket{\phi}$ undergoes dissipation via coupling to the environment, modeled by an ancillary qubit initialized in the ground state $\ket{0}$. (a) Amplitude damping channel, where $\theta$ is obtained from Eq.~\eqref{Eq:angle_AMP}; (b) Excitation channel, where $\theta$ is obtained from Eq.~\eqref{Eq:angle_EXC}; (c) Dephasing channel, where $\theta$ is obtained from Eq.~\eqref{Eq:angle_DEP}; (d) General dissipation channel for the spin-boson model in Eq.~\eqref{eq:HT}. The $\mathrm{R}_y$ rotation angle $\theta$ for each component channel is calculated with $t$ being replaced by the small time step $\tau$ and the damping rates provided in Table~\ref{tab:para_lindblad_SBM}. (e) Real-time evolution of the spin-boson model, where each Trotter layer consists of the evolution unitary $U_\tau = e^{-i H_S \tau }$, followed by the general dissipation channel $\mathcal{E}_{\tau}$.
    }
    \label{fig:noise_circuit_SBM}
\end{figure}
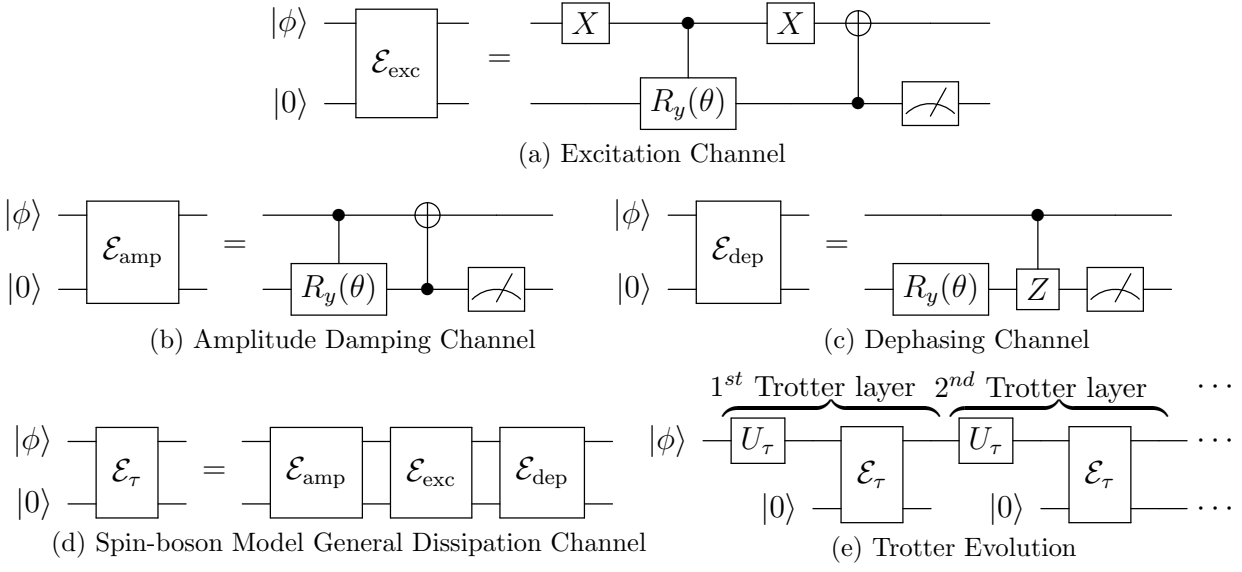

{\em Amplitude Damping Channel:} Consider the amplitude damping channel associated with the $\sigma^+$ jump operator. The corresponding Lindblad equation is:
\begin{equation}\label{Eq:lindblad_ampdamp}
    {{\mathbf{\dot{ \rho}}}}\left( t \right)={\gamma_{\text{amp}}} \left[{\sigma }^{+}\rho (t){\sigma }^{-}-\frac{1}{2}\{{\sigma }^{-}{\sigma }^{+},\rho (t)\}\right],
\end{equation} 
with the initial state
\begin{equation}
    \rho(0) = \left[\begin{array}{cc}
    \rho_{00}(0) & \rho_{01}(0) \\
    \rho_{10}(0) & \rho_{11}(0)
\end{array}\right].
\end{equation}
The analytical solution can then be derived in the form:
\begin{equation}\label{Eq:analy_ampdamp}
    {\rho}(t) =  \begin{bmatrix}
    1-e^{-\gamma_{\text{amp}} t}\rho_{11}(0) & e^{-\frac{\gamma_{\text{amp}}}{2}t}\rho_{01}(0) \\
    e^{-\frac{\gamma_{\text{amp}}}{2}t}\rho_{10}(0) & e^{-\gamma_{\text{amp}} t}\rho_{11}(0)
\end{bmatrix}.
\end{equation}
Alternatively, the amplitude damping channel derived from Eq.~\eqref{eq:isometric_ext_amplitude_damping}, with damping probability $p$, is characterized by the Kraus operators 
$A_0 = \sqrt{p}\ket{0}\bra{1}$ and $A_1=\ket{0}\bra{0} + \sqrt{1-p}\ket{1}\bra{1}$, yielding the analytical solution for the density matrix evolution:
\begin{align} \label{Eq:analy_ampdamp2}
    \rho(t) = \sum_k A_k \rho(0) A^\dagger_k = \left[\begin{array}{cc} 1-(1-p)\rho_{11}(0) & \sqrt{1-p}\rho_{01}(0) \\
    \sqrt{1-p}\rho_{10}(0) & (1-p)\rho_{11}(0)
\end{array}\right] \,.
\end{align}
Comparing Eqs.~\eqref{Eq:analy_ampdamp} and~\eqref{Eq:analy_ampdamp2}, the damping probability $p$ relates to the Lindbladian damping rate $\gamma_{\text{amp}}$ through $1-p=e^{-\gamma_{\text{amp}} t}$, or equivalently, 
\begin{equation} \label{Eq:angle_AMP}
    \cos(\theta_{\text{amp}}/2) = e^{-\gamma_{\text{amp}} t/2},
\end{equation}
allowing the determination of the appropriate rotation angle $\theta_{\text{amp}}$ for the amplitude damping channel $\mathcal{E}_{\rm amp}$ channel shown in Fig.~\ref{fig:noise_circuit_SBM} (b).

{\em Excitation (Inverse Amplitude Damping) Channel:} The excitation process from $\ket{0}$ to $\ket{1}$, corresponding to the $\sigma^-$ jump operator, is similarly described by an amplitude damping channel with the Lindblad equation:
\begin{equation}\label{Eq:lindblad_ampdamp2}
    {{\mathbf{\dot{ \rho}}}}\left( t \right)={\gamma_{\text{exc}}} \left[{\sigma }^{-}\rho (t){\sigma }^{+}-\frac{1}{2}\{{\sigma }^{+}{\sigma }^{-},\rho (t)\}\right].
\end{equation} 
Its quantum circuit implementation, shown in Fig.~\ref{fig:noise_circuit_SBM} (a), is simply an extension of the relaxation channel (Fig.~\ref{fig:noise_circuit_SBM} (b)). Similarly, the Lindbladian excitation rate $\gamma_{\text{exc}}$ relates to the $\mathrm{R}_y$ rotation angle $\theta_{\text{exc}}$ via:
{\allowdisplaybreaks\begin{equation} \label{Eq:angle_EXC}
    \cos\left(\theta_{\text{exc}}/2\right) = e^{-\gamma_{\text{exc}} t/2}.
\end{equation}}

{\em Pure Dephasing Channel:} The dephasing channel associated with the $\sigma^z$ jump operator leads to a decay of off-diagonal coherence elements. The relation between the dephasing rate $\gamma_{\text{dep}}$ and the $\mathrm{R}_y$ rotation angle $\theta_{\text{dep}}$ in the dephasing channel $\mathcal{E}_{\rm dep}$ (Fig.~\ref{fig:noise_circuit_SBM} (c)) is given by:
\begin{equation} \label{Eq:angle_DEP}
    \sin^2\left(\frac{\theta_{\text{dep}}}{2}\right) = \frac{\left(1-e^{-2\gamma_{\text{dep}} t}\right)}{2}.
\end{equation} 
Defining the dephasing probability as $p = \sin^2(\theta/2)$ from Eq.~\eqref{eq:full_unitary_dephasing}, we obtain $1-2p = e^{-2\gamma_{\text{dep}}t}$.

{\em Quantum Circuit for the Spin-Boson Model:} By combining quantum operations from the three dissipation channels, we can construct a quantum circuit that emulates dissipative effects of the spin-boson model for a small time step $\tau$, as shown in Fig.~\ref{fig:noise_circuit_SBM} (d). 
The order of the three different dissipation channels may be important in the general case; however, for small values of $\tau$ (such as in a single Trotter step), the order in which they appear is of less significance.\cite{childs2021theory} We have now arrived at the quantum circuit for evolving the Lindblad equation of the spin-boson model, provided in Fig.~\ref{fig:noise_circuit_SBM} (e). 

\subsection{cQED Modular Hardware Design}\label{ssec:exp}
This section outlines the proposed cQED modular hardware designed to implement our computational framework using available quantum gates.

\begin{figure}[htbp]
    \includegraphics[width = 0.98 \textwidth]{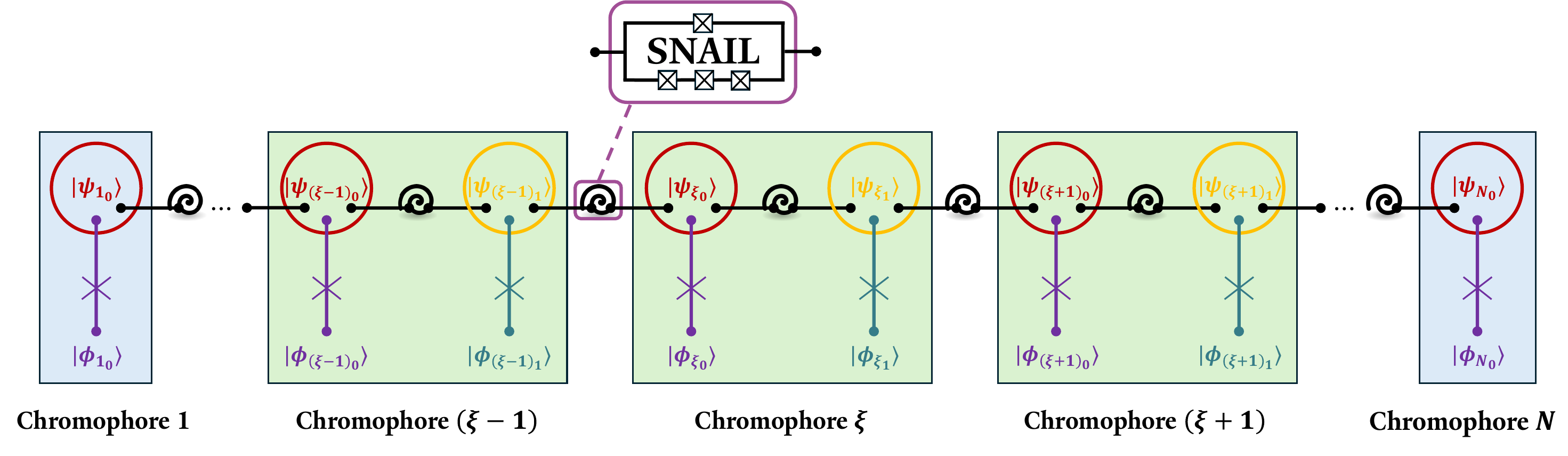}
    \caption{Proposed modular cQED architecture for simulating vibronic dynamics in a 1D molecular chain. Each colored box represents a hardware unit corresponding to a single chromophore. For the two boundary chromophores ($\xi=1, N$), only the high-frequency vibrational modes are considered. Intermediate chromophores $\xi\in[2,N-1]$ are modeled with both high- (red circles) and low-frequency (orange circles) cavities, coupled with SNAILs for efficient cavity-cavity interactions. Transmon qubits (shown in purple) represent the electronic states, while ancillary transmon qubits are depicted in teal blue. 
    }
    \label{fig:modular-cQED}
\end{figure}

The simulation of a 1D array of chromophores is mapped onto a corresponding 1D cQED hardware layout, as shown in Fig.~\ref{fig:modular-cQED}. This architecture employs SNAIL (Superconducting Nonlinear Asymmetric Inductive eLement) couplers to enable efficient coupling mechanisms between resonators.~\cite{Frattini2017SNAIL} Each hardware unit (indicated by a colored box) consists of two cQED devices, including high-frequency (red circle) and low-frequency (yellow circle) modes realized as microwave resonators dispersively coupled to individual superconducting transmon qubits. In this configuration, each chromophore in the 1D chain is mapped to a hardware module, with the full time-evolution decomposed into native operations for the cQED platform.

{\em Instruction Set Architecture (ISA)}: We briefly review the cQED ISA~\cite{liu2024hybrid} employed for simulating vibronic dynamics. In addition to the basic Pauli gates, arbitrary single-qubit rotations can be performed for $\theta \in [0,4\pi)$, 
\begin{align}
    \mathrm{R}_j(\theta) = \exp \left(-i \frac{\theta}{2} \sigma^j\right),
\end{align} 
where $\sigma^j$ represents the Pauli matrices ($j = x, y, z$). This enables native implementation of the Hadamard gate:\begin{align}
    \mathrm{H} = \sigma^x \, \mathrm{R}_y\left(\frac{\pi}{2}\right).
\end{align}

For entangling nearest-neighbor qubits, we utilize the native $XX$-rotation gate, which can be generalized to the $YY$-rotation via single-qubit gate conjugation with $\theta \in [0,4\pi)$:
\begin{equation}
    \mathrm{R}_{XX}(\theta) = \exp\left(-i \frac{\theta}{2} \sigma^x\otimes \sigma^x\right) \label{eq:RXX}
\end{equation}\begin{equation}
    \mathrm{R}_{YY}(\theta) = \exp\left(-i \frac{\theta}{2} \sigma^y\otimes \sigma^y\right) = \left[\mathrm{R}_z\left(-\frac{\pi}{4}\right) \otimes \mathrm{R}_z\left(-\frac{\pi}{4}\right)\right]  \mathrm{R}_{XX}(\theta) \left[\mathrm{R}_z\left(\frac{\pi}{4}\right) \otimes \mathrm{R}_z\left(\frac{\pi}{4}\right)\right]. \label{eq:RYY}
\end{equation}
Additionally, we assume access to Controlled-NOT (CNOT) gates, where non-nearest-neighbor interactions are mediated via nearest-neighbor SWAP operations, each decomposable into three CNOT gates.

For continuous-variable (CV) operations, two fundamental gates are displacement and rotation in the phase-space formulation:~\cite{gerry_knight_2004,Blais2020}  
\begin{align}
    \mathrm{D}(\beta) = \exp\left(\beta b^\dag - \beta^* b\right), \quad\quad
    \mathrm{R}(\theta) = \exp\left(i\theta \hat{n}\right) = \exp\left(i\theta b^\dag b\right), 
\end{align} 
where $b$ and $b^\dag$ are the bosonic annihilation and creation operators, satisfying the canonical commutation relation $[b,b^\dag] = 1$. We note that $\mathrm{R}(\theta)$ rotates the oscillator wavefunction by an angle $\theta \in [0,2\pi)$, while $\mathrm{D}(\beta)$ displaces its Wigner quasiprobability distribution by $\operatorname{Re}{(\beta)}$ along the position axis and $\operatorname{Im}{(\beta)}$ along the momentum axis.

The Fock states $\{\ket{n}\}_{n\in\mathbb{N}}$ are eigenstates of the number operator $\hat{n} = b^\dag b$ and comprise a computational basis: 
\begin{equation}
    \ket{n} = \frac{1}{\sqrt{n!}}\left(b^\dag\right)^n |0\rangle.
\end{equation} 
Non-linear phase-space transformations (also referred to as non-Gaussian operations) enable phase-control over individual Fock states, such as the SNAP gate:~\cite{Krastanov2015}
\begin{align}
    \mathrm{SNAP}(\vec\varphi)&= \sum_{n=0}^\infty e^{-i \varphi_n}|n\rangle\langle n| , \label{eq:SNAPU1}
\end{align}
parameterized by $\vec\varphi=(\varphi_0,\varphi_1,\varphi_2,\ldots,\varphi_{N_\mathrm{max}})$ for $\varphi_n \in [0, 2\pi)$. This gate effectively imparts a different phase to each Fock level of the oscillator. 

For entangling two oscillators, the beam-splitter gate is employed:~\cite{gerry_knight_2004,Zhang_engineering_bilinear_2019,lu2023highfidelity,chapman2022high}
\begin{align}
    \mathrm{BS}(\theta,\varphi)=\exp\left[-i\frac{\theta}{2}\left(e^{i\varphi} b_1^\dag b_2+e^{-i\varphi} b_1b_2^\dag\right)\right],
    \label{eq:BS}
\end{align}
parametrized by the transmittivity $\theta\in[0,4\pi)$ and phase angle $\varphi\in[0,\pi)$.
 
In addition, advances in hybrid continuous-discrete variable (CV-DV) systems have enabled gates that couple the oscillator with its auxiliary qubit in the weakly dispersive regime, like the conditional displacement:~\cite{EickbuschECD,Campagne-Ibarcq2020} 
\begin{align}
    \mathrm{CD}(\beta)
    = \exp\left[\sigma^z\otimes\left(\beta b^\dag - \beta^* b\right)\right], 
    \label{eq:CD}
\end{align}
with $\beta \in \mathbb{C}$. Conditional phase-space rotations are implementable and can be fine-tuned with SNAP gates:~\cite{liu2024hybrid} 
\begin{equation}
    \mathrm{CR(\theta)} = \exp\left[\sigma^z\otimes(i\theta b^\dag b)\right],
\end{equation} for $\theta \in [0,2\pi)$.
 
Universal quantum computation on hybrid CV-DV devices can be realized with the gate set $\{ \text{CD}(\beta), \text{BS}(\theta,\varphi), \text{R}_j(\theta) \}$. For universal oscillator control, the set $\{\text{SNAP}(\vec{\phi}), \text{BS}(\theta,\varphi)$ $,\text{D}(\beta) \}$ suffices.~\cite{liu2024hybrid} Based on these gate sets, the time-evolution governed by the chromophore Hamiltonian in Eq.~\eqref{total-ham-1d} can be efficiently compiled and simulated on cQED devices.

\subsection{Compiling Hamiltonian Simulation with cQED ISA}\label{ssec:ham-sim}
We now describe how to simulate the time-evolution of the system Hamiltonian $\tilde{H}$ from Eq.~\eqref{eq:final-ham-3site}, generalizable to Eq.~\eqref{total-ham-1d}. In this framework, $H_2^{(\zeta)}$ in the rotated frame is decomposed into $\tilde{H}_{2,XX}^{(\zeta)}$ and $\tilde{H}_{2,YY}^{(\zeta)}$, as outlined in Appendix~\ref{asec:H-derivation} for $N=3$ chromophores. This decomposition implements only the cQED ISA described in Sec.~\ref{ssec:exp}. 

Given a discrete time step $\tau$, the objective is to compute the time evolution of the system at each time $t$: 
\begin{equation}
    \ket{\Psi(t+\tau)} = \exp\left(-\frac{i}{\hbar}\tilde{H}\tau\right)\ket{\Psi(t)},
\end{equation} where $\ket{\Psi(t)}$ denotes the full state-vector of the $N$-site chromophore system at time $t$. In our notation, the $\xi^\text{th}$ chromophore consists of a pair of electronic states with high and low frequency, represented as qubits $\ket{\phi_{\xi_{0}}}$ and $\ket{\phi_{\xi_{1}}}$, respectively. Each electronic state is coupled to an associated vibrational mode, encoded as a qumode in states $\ket{\psi_{\xi_{0}}}$ and $\ket{\psi_{\xi_{1}}}$. 

For the decomposition of $e^{-i \tilde{H} \tau/\hbar}$ into elementary gates suitable for cQED implementation, we leverage established techniques from Hamiltonian simulation. These include gate decompositions for (i) qubit-centric systems, such as the Heisenberg spin chain model~\cite{salathe2015digital,Oftelie2022Sim} and Kitaev’s honeycomb model,~\cite{Raeisi2012Sim} where interactions are mapped onto sequences of single- and two-qubit gates; (ii) qumode-centric systems including the multi-site Bose–Hubbard model;~\cite{Kalajdzievski2018BoseHubbard} and (iii) hybrid qubit-qumode systems, notably recent developments in the simulation of gauge fields.~\cite{crane2024hybrid} These prior works provide the foundational strategies for gate decomposition applied to the unique structure of our chromophore model.

\subsubsection{Real-time Evolution via Trotterization} \label{sssec:trotter}
To simulate the real-time evolution of the system, we employ a Trotter-Suzuki decomposition, which enables the approximation of the time-evolution operator by sequentially applying exponentials of Hamiltonian terms that are natively implementable on quantum hardware. The key challenge lies in properly decomposing the total Hamiltonian $\tilde{H}$ into separate terms, each compatible with available operations on the cQED platform. However, these terms generally do not commute, so the Baker–Campbell–Hausdorff formula for Trotterization introduces errors. 

We refer the readers to the established error analysis of Trotterization,\cite{childs2021theory} including recent extensions to bosonic devices.~\cite{kang2023leveraging} While the Hamiltonian decomposition can, in principle, be optimized to minimize Trotter error, practical hardware constraints often impose limitations. For example, terms such as $\frac{g_{ab}}{2}(\sigma_a^x\sigma_b^x)$ and $\frac{g_{ab}}{2}(\sigma_a^x\sigma_b^x)(l+l^\dag)$ in Eq.~\eqref{eq:final-ham-3site-h2xx} cannot currently be implemented as a single native operation. Instead, they must be decomposed into separate operations, increasing Trotter error. This trade-off between leveraging native hardware capabilities and minimizing Trotterization error is a key consideration in quantum simulation.

{\em Hamiltonian Decomposition and Trotterization Scheme}: For our system, we rewrite the total Hamiltonian from Eq.~\eqref{eq:final-ham-3site} as a sum of four distinct terms:
\begin{equation}
    \tilde{H} = \tilde{H}_0 + \tilde{H}_1 + \tilde{H}_{2,XX} + \tilde{H}_{2,YY}.
\end{equation} This can be reorganized as
\begin{equation}
    \tilde{H} = \underbrace{(1-w_1-2w_2)\tilde{H}_0}_{\mathcal{H}_0} + \underbrace{(w_1\tilde{H}_0+\tilde{H}_1)}_{\mathcal{H}_1} + \underbrace{(w_2\tilde{H}_0 + \tilde{H}_{2,XX})}_{\mathcal{H}_2} + \underbrace{(w_2\tilde{H}_0+\tilde{H}_{2,YY})}_{\mathcal{H}_3},
\end{equation}
where $w_1 \in [0, 1]$ and $w_2 \in [0, 1/2]$ are tunable weights used to distribute the free evolution term $\tilde{H}_0$ across different Trotter steps. Each term represents different physical interactions and demands distinct implementation strategies. The terms in $\tilde{H}_0$ can be toggled on or off at will during the simulation on cQED devices while the terms in $\tilde{H}_1$ can only be turned on and off simultaneously in an analog manner. $\tilde{H}_{2, XX}$ and $\tilde{H}_{2, YY}$ pose the greatest challenge, as they are not directly implementable on cQED hardware and must be synthesized/compiled from native gates.

To simulate the time evolution over a small time step $\tau$, we apply the second-order Suzuki-Trotter formula:~\cite{suzuki1991general}
\begin{align}
    e^{-\frac{i}{\hbar}\tilde{H} \tau} 
    \approx & \prod_{p=0}^3 e^{-\frac{i}{\hbar}\mathcal{H}_p\tau/2} \prod_{q=0}^3 e^{-\frac{i}{\hbar}\mathcal{H}_q\tau/2} + O(\alpha_{\rm comm}\tau^3),
    \label{eq:2nd-trotter}
\end{align}
where the leading-order error term arises from the non-commutativity of the Hamiltonian components. The Trotter error coefficient, $\alpha_{\rm comm}$, quantifies this error and is given by:~\cite{childs2021theory}
\begin{align}
    & \alpha_{\rm comm} 
    = \frac{1}{12} \Bigl\{ \| \left[\mathcal{H}_1+\mathcal{H}_2+\mathcal{H}_3,[\mathcal{H}_1+\mathcal{H}_2+\mathcal{H}_3,\mathcal{H}_0]\right] \nonumber \\& + \| \left[\mathcal{H}_2+\mathcal{H}_3,[\mathcal{H}_2+\mathcal{H}_3,\mathcal{H}_1]\right] \| + \| \left[\mathcal{H}_3,[\mathcal{H}_3,\mathcal{H}_2]\right] \| \Bigl\} \nonumber \\ & + \frac{1}{24} \Bigl\{ \| \left[\mathcal{H}_0,[\mathcal{H}_0,\mathcal{H}_1+\mathcal{H}_2+\mathcal{H}_3]\right] \| + \| \left[\mathcal{H}_1,[\mathcal{H}_1,\mathcal{H}_2+\mathcal{H}_3] \right] \| + \| \left[ \mathcal{H}_2, [\mathcal{H}_2,\mathcal{H}_3]\right] \| \Bigl\},
\end{align}
where $\|\cdot\|$ denotes the spectral norm. This expression captures the dominant error contributions arising from nested commutators of the Hamiltonian components.

{\em Error Mitigation and Parameter Selection}: In principle, the weights $w_1$ and $w_2$ can be optimized to minimize $\alpha_{\rm comm}$ and thereby reduce Trotter errors. However, for the purpose of this work, we focus on high-accuracy simulations by setting $w_1 = w_2 = 0$, effectively simplifying the decomposition. To ensure the Trotter error remains negligible, we select a sufficiently small time step $\tau$ such that $\alpha_{\rm comm} \tau^2 \ll 1$, or equivalently,
\begin{align}
    \tau \ll \frac{1}{\sqrt{\alpha_{\rm comm}}}. 
\end{align}
This condition guarantees that the accumulated error over the simulation remains reasonably bounded, balancing computational efficiency with the desired accuracy.

\subsubsection{Compiling Quantum Circuits to Simulate Dispersive Vibronic Couplings} \label{sssec:compiled_circuit_trotter}
To simulate the generalized multi-site Hamiltonian in Eq.~\eqref{total-ham-1d}, we compile each term into its quantum circuit native implementation. In this subsection, we only focus on the terms
\begin{align}
\frac{g_{\xi_0, (\xi\pm1)_0, \xi_1}}{2} (\sigma^+_{\xi_0} \sigma^-_{(\xi\pm1)_0}) (b_{\xi_1} + b^\dagger_{\xi_1}),   
\end{align}
which describe dispersive vibrational-electronic coupling between adjacent chromophores. The readers are referred to Appendix \ref{asec:compiled_circuit_trotter} for the full compilation of the remaining terms in Eq.~\eqref{total-ham-1d}. Following Appendix~\ref{asec:H-derivation}, the simulation of $\sigma^+\sigma^-$ interactions is split into separate $\sigma^x\sigma^x$- and $\sigma^y\sigma^y$-interaction terms, compiled via Trotterization with parametrized angles
\begin{align}
    \theta = \frac{g_{\xi_0, (\xi\pm1)_0, \xi_1}\tau}{2},
\end{align}
for the $XX$- and $YY$-rotations.

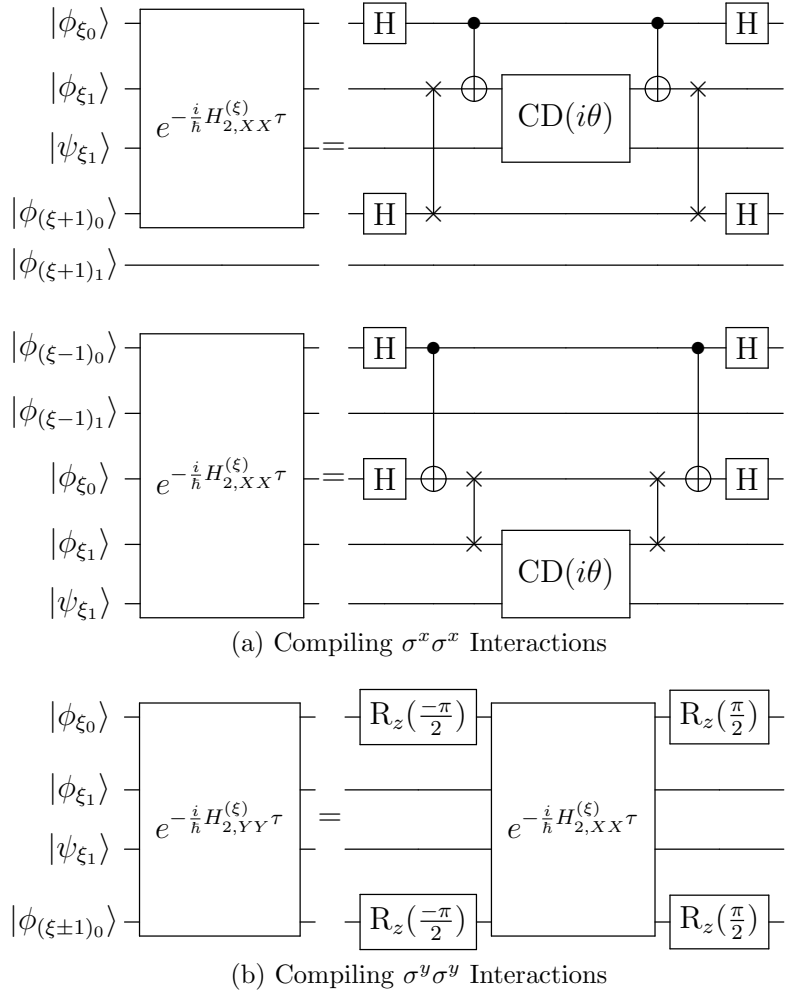
\begin{figure}[htbp]
    \centering
    \subfloat[Compiling $\sigma^x\sigma^x$ Interactions]{
        \mbox{
            \Qcircuit @C=0.48em @R=1em {
            & \ket{\phi_{\xi_0}} &&&& \multigate{3}{e^{-\frac{i}{\hbar}H^{(\xi)}_{2,XX}\tau}} & \qw & 
                & \quad & \gate{\mathrm{H}} & \qw & \ctrl{1} & \qw & \ctrl{1} & \qw & \gate{\mathrm{H}} & \qw \\
            & \ket{\phi_{\xi_1}} &&&& \ghost{e^{-\frac{i}{\hbar}H^{(\xi)}_{2,XX}\tau}} & \qw & 
                & \quad & \qw & \qswap & \targ & \multigate{1}{\text{CD}(i\theta)} & \targ & \qswap & \qw & \qw \\
            & \ket{\psi_{\xi_1}} &&&& \oghost{e^{-\frac{i}{\hbar}H^{(\xi)}_{2,XX}\tau}} & \ow & =
                & \quad & \ow & \ow \qwx & \ow & \oghost{\text{CD}(i\theta)} & \ow & \ow \qwx & \ow & \ow \\
            & \ket{\phi_ {(\xi+1)_0}} \quad &&&& \ghost{e^{-\frac{i}{\hbar}H^{(\xi)}_{2,XX}\tau}} & \qw & 
                & \quad & \gate{\mathrm{H}} & \qswap \qwx & \qw & \qw & \qw & \qswap \qwx & \gate{\mathrm{H}} & \qw \\
            & \ket{\phi_{(\xi+1)_1}} \quad &&&& \qw & \qw & 
                & \quad & \qw & \qw & \qw & \qw & \qw & \qw & \qw & \qw \\ \\
            & \ket{\phi_ {(\xi-1)_0}} \quad &&&& \multigate{4}{e^{-\frac{i}{\hbar}H^{(\xi)}_{2,XX}\tau}} & \qw &
                & \quad & \gate{\mathrm{H}} & \ctrl{2} & \qw & \qw & \qw & \ctrl{2} & \gate{\mathrm{H}} & \qw \\
            & \ket{\phi_{(\xi-1)_1}} \quad &&&& \ghost{e^{-\frac{i}{\hbar}H^{(\xi)}_{2,XX}\tau}} & \qw & 
                & \quad & \qw & \qw & \qw & \qw & \qw & \qw & \qw & \qw \\
            & \ket{\phi_{\xi_0}} &&&& \ghost{e^{-\frac{i}{\hbar}H^{(\xi)}_{2,XX}\tau}} & \qw & =
                & \quad & \gate{\mathrm{H}} & \targ & \qswap & \qw & \qswap & \targ & \gate{\mathrm{H}} & \qw \\
            & \ket{\phi_{\xi_1}} &&&& \ghost{e^{-\frac{i}{\hbar}H^{(\xi)}_{2,XX}\tau}} & \qw & 
                & \quad & \qw & \qw & \qswap \qwx & \multigate{1}{\text{CD}(i\theta)} & \qswap \qwx & \qw & \qw & \qw \\
            & \ket{\psi_{\xi_1}} &&&& \oghost{e^{-\frac{i}{\hbar}H^{(\xi)}_{2,XX}\tau}} & \ow & 
                & \quad & \ow & \ow & \ow & \oghost{\text{CD}(i\theta)} & \ow & \ow & \ow & \ow
            }
        }
    } \\
    \subfloat[Compiling $\sigma^y\sigma^y$ Interactions]{
        \mbox{
            \Qcircuit @C=0.48em @R=1em {
            &\ket{\phi_{\xi_0}} &&&& \multigate{3}{e^{-\frac{i}{\hbar}H^{(\xi)}_{2,YY}\tau}} & \qw & 
                & \quad & \gate{\mathrm{R}_z(\frac{-\pi}{2})} & \multigate{3}{e^{-\frac{i}{\hbar}H^{(\xi)}_{2,XX}\tau}} & \gate{\mathrm{R}_z(\frac{\pi}{2})} & \qw \\
            &\ket{\phi_{\xi_1}} &&&& \ghost{e^{-\frac{i}{\hbar}H^{(\xi)}_{2,YY}\tau}} & \qw & \raisebox{-2em}{=}
                & \quad & \qw & \ghost{e^{-\frac{i}{\hbar}H^{(\xi)}_{2,XX}\tau}} & \qw & \qw \\
            &\ket{\psi_{\xi_1}} &&&& \oghost{e^{-\frac{i}{\hbar}H^{(\xi)}_{2,YY}\tau}} & \ow & 
                & \quad & \ow & \oghost{e^{-\frac{i}{\hbar}H^{(\xi)}_{2,XX}\tau}} & \ow & \ow \\
            &\ket{\phi_{(\xi\pm1)_0}} \quad &&&& \ghost{e^{-\frac{i}{\hbar}H^{(\xi)}_{2,YY}\tau}} & \qw & 
                & \quad & \gate{\mathrm{R}_z(\frac{-\pi}{2})} & \ghost{e^{-\frac{i}{\hbar}H^{(\xi)}_{2,XX}\tau}} & \gate{\mathrm{R}_z(\frac{\pi}{2})} & \qw
            }
        }
    }
    \caption{(a) Two circuit compilations for $\sigma^x\sigma^x$ interaction terms between adjacent chromophores. These circuits reduced to qubit operations and transmon-cavity dispersive interactions on the low-frequency mode $\xi_1$. The second circuit implicitly requires a pair of conjugate SWAP operations to mediate non-nearest-neighbor CNOT gates. (b) Circuit compilation for $\sigma^y\sigma^y$ interaction terms between adjacent chromophores. The $H_{2,XX}$ block is implemented as shown in (a). For the cQED hardware layout in Fig.~\ref{fig:modular-cQED}, the $(\xi-1)_0$ qubit shall be placed before the $\xi_0$ qubit.}
    \label{fig:two-hard-qubit-soft-mode}
\end{figure}

{\em Compiling $\sigma^x\sigma^x$ Interactions}: To simulate the $\sigma^x\sigma^x$-interaction terms, we conjugate a conditional displacement operation with CNOT and SWAP gates, yielding
\begin{align}
    &e^{i\theta \sigma^x_{\xi_0} \sigma^x_{(\xi + 1)_0} (b_{\xi_1} + b^\dagger_{\xi_1})} \nn \\
    =& (\mathrm{H}_{\xi_0} \mathrm{H}_{(\xi + 1)_0}) \swap_{{(\xi + 1)_0}{\xi_1}} 
        \Bigr[ \cnot{{\xi_0}}{\xi_1} e^{i\theta \sigma^z_{\xi_1} (b_{\xi_1} + b^\dagger_{\xi_1})} \cnot{\xi_0}{\xi_1} \Bigr] 
           \swap_{{(\xi + 1)_0}{\xi_1}} (\mathrm{H}_{\xi_0} \mathrm{H}_{(\xi + 1)_0}) \nn \\
    =& (\mathrm{H}_{\xi_0} \mathrm{H}_{(\xi + 1)_0}) \swap_{{(\xi + 1)_0}{\xi_1}} 
    \Bigr[ \cnot{{\xi_0}}{\xi_1} \mathrm{C}_{\xi_1}\mathrm{D}_{\xi_1}(i\theta)  \cnot{\xi_0}{\xi_1} \Bigr] \swap_{{(\xi + 1)_0}{\xi_1}} (\mathrm{H}_{\xi_0} \mathrm{H}_{(\xi + 1)_0}), \label{two-transmon-one-cavity-a}
\end{align} 
alongside an alternative decomposition, as shown in Fig.~\ref{fig:two-hard-qubit-soft-mode} (a):
{\allowdisplaybreaks\begin{align}
    &e^{i\theta \sigma^x_{\xi_0} \sigma^x_{(\xi - 1)_0} (b_{\xi_1} + b^\dagger_{\xi_1})} \nn \\
    =& (\mathrm{H}_{\xi_0} \mathrm{H}_{(\xi - 1)_0}) \cnot{{(\xi - 1)_0}}{\xi_0} 
        \Bigr[ \swap_{{\xi_0}{\xi_1}} e^{i\theta \sigma^z_{\xi_1} (b_{\xi_1} + b^\dagger_{\xi_1})} \swap_{{\xi_0}{\xi_1}} \Bigr] 
            \cnot{{(\xi - 1)_0}}{\xi_0} (\mathrm{H}_{\xi_0} \mathrm{H}_{(\xi - 1)_0}), \nn \\
    =& (\mathrm{H}_{\xi_0} \mathrm{H}_{(\xi - 1)_0}) \cnot{{(\xi - 1)_0}}{\xi_0} 
        \Bigr[ \swap_{{\xi_0}{\xi_1}} \mathrm{C}_{\xi_1}\mathrm{D}_{\xi_1}(i\theta) \swap_{{\xi_0}{\xi_1}} \Bigr] 
            \cnot{{(\xi - 1)_0}}{\xi_0} (\mathrm{H}_{\xi_0} \mathrm{H}_{(\xi - 1)_0}). \label{two-transmon-one-cavity-b}
\end{align}}
In Eqs.~\eqref{two-transmon-one-cavity-a} and Eq.~\eqref{two-transmon-one-cavity-b}, $\mathrm{H}$ denotes the Hadamard gate, and the SWAP gates mediate interactions between non-nearest-neighbor qubits in the cQED architecture. Eq.~\eqref{two-transmon-one-cavity-a} describes vibronic interactions with the {\em next} chromophore in the array, requiring nearest neighbor SNAIL couplings between $\xi_0-\xi_1$ and  $\xi_1-(\xi+1)_1$. Eq.~\eqref{two-transmon-one-cavity-b} describes interactions with the {\em previous} chromophore, requiring three mediations: $(\xi-1)_0-(\xi-1)_1$, $(\xi-1)_1-\xi_0$, and $\xi_0-\xi_1$.

{\em Compiling $\sigma^y\sigma^y$ Interactions}: The $\sigma^y\sigma^y$-interaction terms can also be simulated in a very similar manner to $\sigma^x\sigma^x$-interaction terms, using the identity 
\begin{align}
e^{-i\frac{\pi}{4}\sigma^z} \sigma^x e^{i\frac{\pi}{4}\sigma^z} = \sigma^y,    
\end{align}
which implies
\begin{align}
    & e^{i\theta \sigma^y_{\xi_0} \sigma^y_{(\xi\pm1)_0} ({(\xi\pm1)_0})}= \left(e^{i\frac{\pi}{4}\sigma^z_{\xi_0}} \otimes e^{i\frac{\pi}{4}\sigma^z_{(\xi\pm1)_0}}\right) \left[ e^{i\theta \sigma^x_{\xi_0} \sigma^x_{(\xi\pm1)_0} (b_{\xi_1} + b^\dagger_{\xi_1})}\right] \left(e^{i\frac{\pi}{4}\sigma^z_{\xi_0}}\otimes e^{i\frac{\pi}{4}\sigma^z_{(\xi\pm1)_0}}\right).
\end{align}
This results in the circuit as shown in Fig.~\ref{fig:two-hard-qubit-soft-mode} (b).

\subsubsection{Resource Estimation from Hybrid ISA}
\label{ssec:resource-est}
In this section, we estimate the resources required to simulate the Trotterized 3-site chromophore model. Specifically, we count two-qubit and qubit-qumode gates based on the hybrid cQED ISA.

The first three terms of Eq.~\eqref{eq:final-ham-3site-h1} require 3 SNAP gates, while the remaining three terms need 3 CD operations. For the circuit shown in Fig.~\ref{fig:compile_H} (d), each SWAP operation decomposes into three CNOT gates, impying that a single transmon-transmon coupling requires 7 nearest-neighbor two-qubit gates. The four two-transmon interaction terms in Eqs.~\eqref{eq:final-ham-3site-h2xx}-\eqref{eq:final-ham-3site-h2yy} cumulatively demand 28 nearest-neighbor two-qubit gates. 

For the circuit in Fig.~\ref{fig:compile_H} (c), the gate requirements are equivalent to one CD gate and 6 nearest-neighbor CNOT gates. Therefore, the two $\sigma^z(l+l^\dagger)$ terms in Eqs.~\eqref{eq:final-ham-3site-h2xx}-\eqref{eq:final-ham-3site-h2yy} require 2 CD operations and 12 CNOT gates.

In Fig.~\ref{fig:two-hard-qubit-soft-mode} (a), the first circuit requires 1 CD gate, 2 CNOT gates for entangling transmons $a$ and $l$, and 6 CNOT gates for SWAP operations between transmons $b$ and $l$. The second circuit requires 1 CD gate, 6 CNOT gates for the two SWAP operations between transmons $a$ and $l$, plus 8 CNOT gates to account for the two CNOT operations between transmons $a$ and $b$. The latter gates, separated by the low-frequency cavity coupled to $b$, require two additional nearest-neighbor SWAP operations. 

Cosidering both $\sigma^x\sigma^x(l + l^\dagger)$ (Fig.~\ref{fig:two-hard-qubit-soft-mode} (a)) and $\sigma^y\sigma^y(l+l^\dagger)$ terms (Fig.~\ref{fig:two-hard-qubit-soft-mode} (b)), the gate count amounts to 2 CD and 28 CNOT gates for $a$-$c$ interactions, and 2 CD and 16 CNOT gates for $a$-$b$ interactions. This results in a total of 4 CD and 44 CNOT gates for two-transmon-one-cavity operations. 

Summing the contributions, the 3-site chromophore model requires per Trotter step: 84 CNOT gates, 9 CD gates, and 3 SNAP gates. For a generalizing 1D array of $N$-chromophores, assuming negligible low-frequency modes at the boundaries and mapping to $2N-2$ transmon qubits and $2N-2$ cavities, the total gate count per Trotter step is \begin{align}
    N_{\rm gate} = (N-2) \times (84 ~\text{CNOT} + 9 ~\text{CD} + 3 ~\text{SNAP}). \label{eq:gate_count}
\end{align}

{\em Cavity-Only Architecture}: In an alternative scenario where transmon connectivity is absent, we consider a cavity-only approach. Here, we assume native access to CD operations via weak dispersive interactions between each cavity and its coupled transmon. Each CNOT gate can be analytically decomposed into four native beam-splitter (BS) gates between adjacent cavities and four CD operations.~\cite{liu2024hybrid} Consequently, simulating the 3-site chromophore model requires 336 BS gates, 345 CD gates, and 3 SNAP gates per Trotter step. Extending this to an $N$ chromophore 1D array, the total gate count per Trotter step is
\begin{equation}
    N_{\rm gate} = (N-2) \times (336~ \text{BS} + 345 ~\text{CD} + 3 ~\text{SNAP}). \label{eq:alt_gate_count}
\end{equation}

\section{Results} \label{sec:results}
\subsection{Validation Against Exact Lindbladian Dynamics:} To assess the accuracy of the proposed quantum circuits in capturing environmental effects (Sec.~\ref{ssec:engineering-dissipation}), we compare the simulation results with exact Lindblad dynamics for the spin-boson model. Specifically, we consider a Debye spectral density: 
\begin{equation}
    J(\omega) = \frac{\eta \omega \omega_c}{\omega^2 + \omega_c^2},
\end{equation}
using parameters representative of photoinduced charge transfer in solution:\cite{sun18,tong20,dan25} system-bath coupling strength $\eta = 0.3$~eV, spectral width $\omega_c = 30~{\rm cm}^{-1}$, site energy $E_0 = 0.2$~eV, and temperature $T= 77$~K. The environmental coupling is assumed to be equally distributed among all Pauli operators ($\eta_x=\eta_y=\eta_z = 1/3$), and the system is initialized in the superposition state $\rho(0) = |+\rangle\langle +|$. 

\begin{figure}[htbp]
    \centering
    \includegraphics[width=0.6\textwidth]{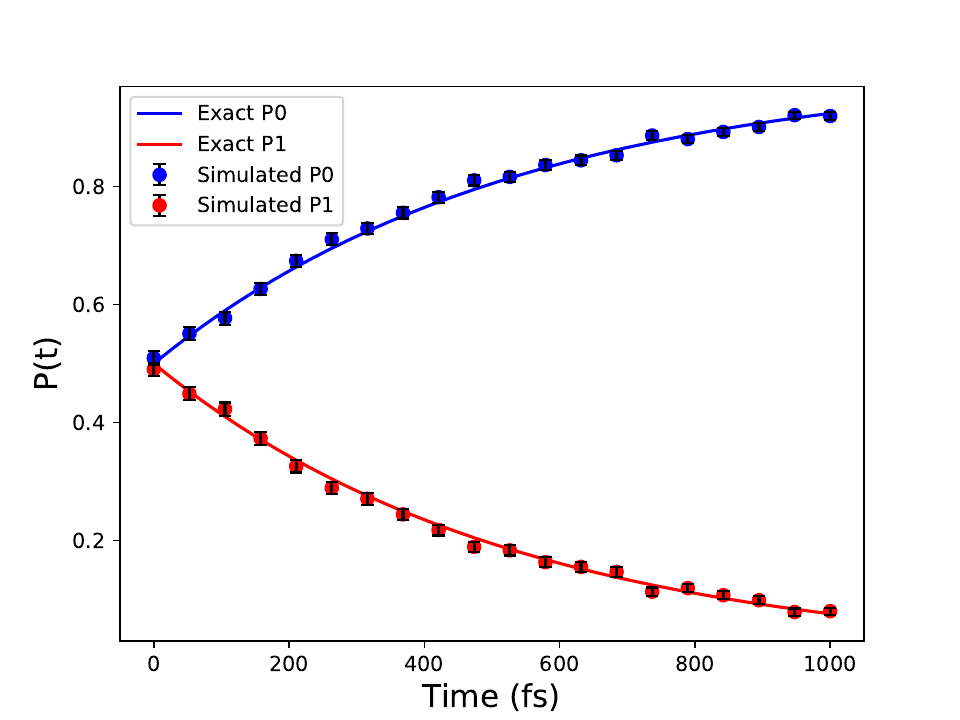}
    \caption{ Population dynamics of the spin-boson model. The results compare Lindblad dynamics simulated using {\scshape QuTiP} with the Trotterized quantum circuit in Fig.~\ref{fig:noise_circuit_SBM} (e). $P_0$ and $P_1$ denote the probabilities of measuring $\ket{0}$ and $\ket{1}$, respectively, for the system qubit. Each data point represents the average measurement from 2000 shots.}
    \label{fig:lindblad_simulations}
\end{figure}

Fig.~\ref{fig:lindblad_simulations} compares the population dynamics obtained from the quantum circuit simulations using {\scshape AerSimulator} (from {\scshape Qiskit Aer})~\cite{qiskit24} to those obtained with the exact Lindblad dynamics computed the {\scshape QuTiP} solver.~\cite{johansson12,johansson13} This result confirms that the circuit in Fig.~\ref{fig:noise_circuit_SBM} (e) accurately captures the general characteristics of environmentally induced dissipative effects within the validity regime of the Lindblad formalism.

\subsection{Non-dissipative Simulations}
To assess the accuracy of the compiled quantum topology introduced in Sec.~\ref{ssec:ham-sim}, we benchmarked its performance by propagating the system Hamiltonian (Eq.~\eqref{eq:final-ham-3site}) using the numerical solver method implemented in {\scshape QuTiP}.~\cite{johansson12,johansson13} This corresponds to numerically solving the \textit{Liouvillian} part of the Lindblad equation in the absence of dissipation, {\em i.e.}, all damping rates are set to zero ($\gamma_j = 0$).
\begin{figure}
    \centering
    \includegraphics[width=0.6\linewidth]{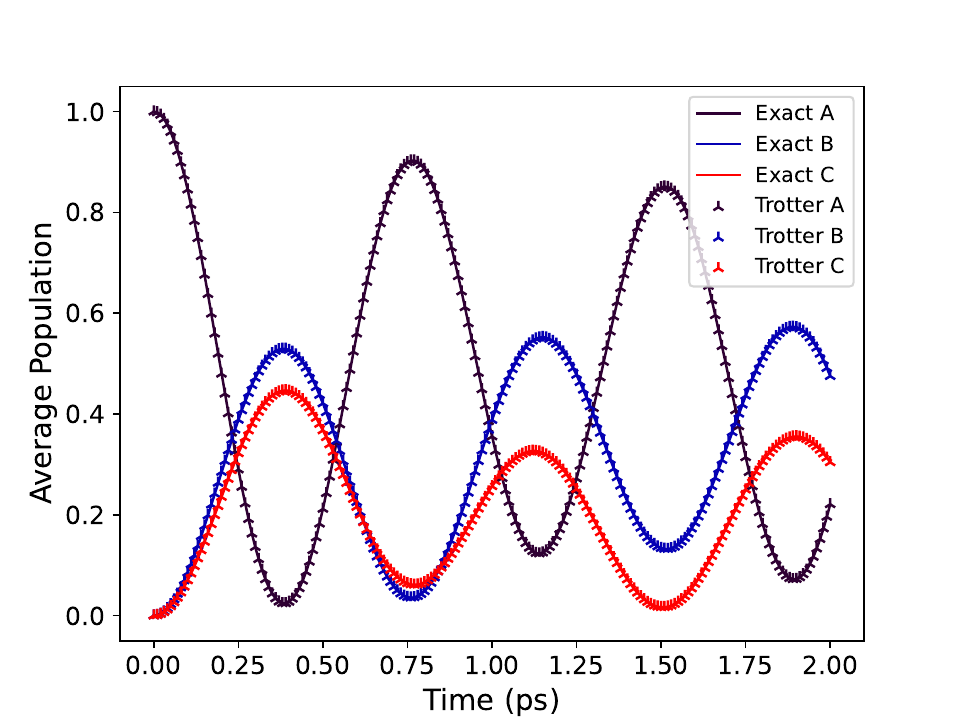}
    \caption{Population dynamics of the 3-site chromophore system without dissipation over a 2 picosecond (ps) timescale, comparing exact evolution computed with {\scshape QuTiP} (solid lines) and Trotterized quantum simulation using {\scshape Bosonic-Qiskit}~\cite{Biskit} (markers). Each data point represents the average measurement from 10,000 shots, with a Fock space truncated to 8 levels applied in both simulations.}
    \label{fig:exact_simulation}
\end{figure}

Given the presence of both electronic and vibrational transitions in the chromophore system, we utilized {\scshape Bosonic-Qiskit}\cite{Biskit}, an extension of {\scshape Qiskit} that enables Trotterized simulations of hybrid CV-DV systems via {\scshape Qiskit Aer} simulators. Fig.~\ref{fig:exact_simulation} compares the exact quantum evolution with the Trotterized quantum simulation, where each time step corresponds to approximately 10 femtoseconds (fs). The close agreement between both simulations validates the accuracy of our approach. 

\subsection{Dissipative Simulations}
Dissipation is a fundamental aspect of real-world quantum systems and must be incorporated into physically relevant simulations. Here, we combine amplitude damping and dephasing channels to effectively capture key features of environmentally induced dissipation in the 3-site chromophore model system. 
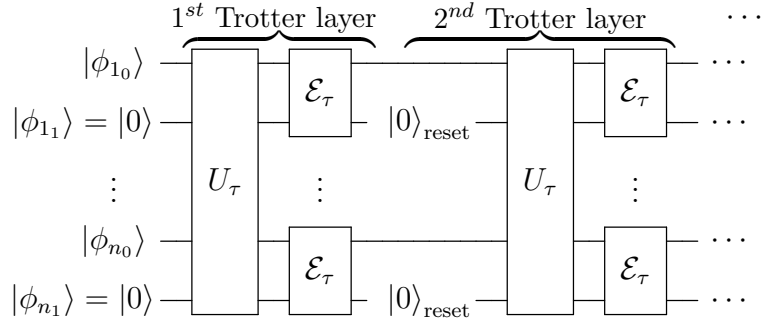
\begin{figure}[htbp]
            \begin{center}
            \mbox{
                \Qcircuit @C=0.5em @R=1.0em {
                    &&&&&& \text{\small $1^{st}$ Trotter layer} &&&&&&&&&&& \text{\small $2^{nd}$ Trotter layer} &&&&&&& \cdots \\
                    \ket{\phi_{1_0}} &&&& \qw & \multigate{4}{U_\tau} 
                        & \qw & \multigate{1}{\mathcal{E}_\tau} & \qw & \qw & \qw & \qw & \qw & \qw & \qw & \qw & \qw & \multigate{4}{U_\tau} 
                            & \qw & \multigate{1}{\mathcal{E}_\tau} & \qw & \qw && \cdots \\
                    \ket{\phi_{1_1}}=\ket{0} \quad \quad &&&& \qw & \ghost{U_\tau} 
                        & \qw & \ghost{\mathcal{E}_\tau} & \qw &&&& \ket{0}_{\text{reset}} &&&& \qw & \ghost{U_\tau} 
                             & \qw & \ghost{\mathcal{E}_\tau} & \qw & \qw && \cdots\\
                    \vdots &&&&& \nghost{U_\tau} 
                        && \vdots &&&&&&&&&& \nghost{U_\tau} 
                            && \vdots &&  \\
                    \ket{\phi_{n_0}} &&&& \qw & \ghost{U_\tau} 
                        & \qw & \multigate{1}{\mathcal{E}_\tau} & \qw & \qw & \qw & \qw & \qw & \qw & \qw & \qw & \qw & \ghost{U_\tau} 
                             & \qw & \multigate{1}{\mathcal{E}_\tau} & \qw & \qw && \cdots \\
                    \ket{\phi_{n_1}}=\ket{0} \quad \quad &&&& \qw & \ghost{U_\tau} 
                        & \qw & \ghost{\mathcal{E}_\tau} & \qw &&&& \ket{0}_{\text{reset}} &&&& \qw & \ghost{U_\tau} 
                             & \qw & \ghost{\mathcal{E}_\tau} & \qw & \qw && \cdots
                    \gategroup{2}{6}{5}{9}{.5em}{^\}}
                    \gategroup{2}{12}{5}{20}{.5em}{^\}}
                }
            }
            \end{center}
            \caption{Generalized quantum circuit topology for simulating a dissipative 1D-array of $n$ chromophores. In each Trotter step $\tau$, the full system Hamiltonian from Eq.~\eqref{total-ham-1d} is first propagated, followed by the quantum dissipative channels $\mathcal{E}_\tau$, as in Fig.~\ref{fig:noise_circuit_SBM} (d), to the low-frequency qubits $\ket{\phi_{\xi_1}}$. The symbols $\ket{0}_{\text{reset}}$ indicate that these qubits are then incoherently reset to $\ket{0}$ state after each dissipation step, independent of measurement outcomes.}
            \label{fig:circuit-trotter-damping}
\end{figure}

To emulate quantum dissipative channels, we implement a gate-based approach following Ref.~\citenum{Guimar_es_2023} where we measure the ancilla qubits and reset them to the ground state after each Trotter step (Sec.~\ref{sssec:compiled_circuit_trotter}). The overall structure of this approach is illustrated in Fig.~\ref{fig:circuit-trotter-damping}, where low-frequency qubits $\ket{\phi_{\xi_1}}$ are used to implement the dissipative channels. These qubits serve to control the evolution of the low-frequency qumodes $\ket{\psi_{\xi_1}}$ rather than evolving in real-time themselves (see compiled circuits in Fig.~\ref{fig:noise_circuit_SBM}). These channels are parametrized in terms of the dissipative Lindbladian damping rates and associated jump operators of the system, as discussed in Sec.~\ref{sec:ehcqed}.

Given a Trotter step of duration $\tau$, the damping rates for the amplitude damping and dephasing channels are given by $\gamma_{\rm amp}\tau$ and $\gamma_{\rm dep}\tau$, respectively. The corresponding $\mathrm{R}_y$ rotation angles for these dissipative channels are determined by:
\begin{align}
    \theta_{\rm amp} &= 2\arcsin{\sqrt{\gamma_{\rm amp}\tau}}, \\
    \theta_{\rm dep} &= 2\arcsin{\sqrt{\gamma_{\rm dep}\tau}},
\end{align} 
where $\gamma_{\rm amp}$ and $\gamma_{\rm dep}$ are the damping rates.  Further details on performance and convergence analysis, including the choice of Trotter step size $\tau=$ 10 fs, a Fock truncation of 8 levels, and 10,000 shots per simulation, are provided in Appendix \ref{asec:Performance-Convergence}.

To investigate the impact of environmental dissipation on the 3-site chromophore system, we analyze population dynamics under varying amplitude damping and dephasing rates. These simulations help elucidate how energy and quantum coherence evolve in open quantum systems and provide insight into how environmental effects can be tuned to control energy transfer pathways. 

\begin{figure}[H]
    \centering
    \includegraphics[width=0.6 \textwidth]{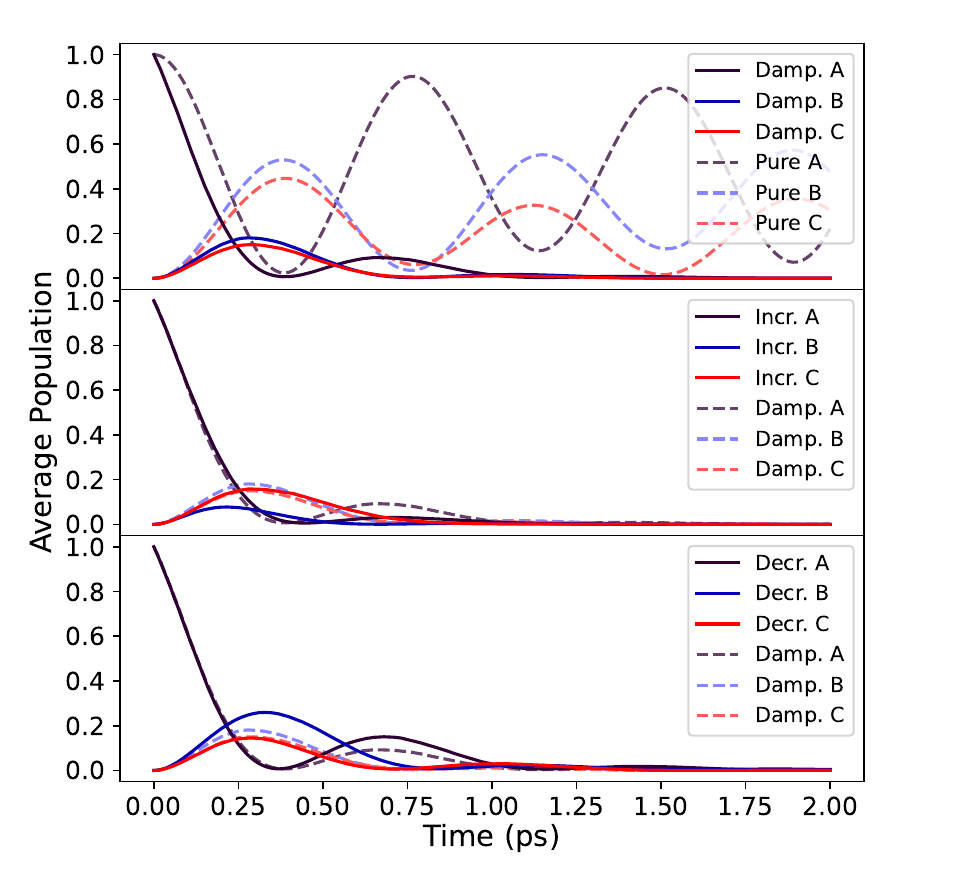}
    \caption{Population dynamics of the 3-site chromophore system under various damping rates. The top graph shows the population dynamics of the 3-site chromophore ($\gamma_{\rm amp,all}=3.15$ THz) under amplitude damping, plotted against a non-dissipative system. The middle and bottom graphs demonstrate the effects of tuning damping dissipation on chromophore $B$, with the middle graph showing the effects of a 3$\times$ increase ($\gamma_{\rm amp,b} = 9.45$ THz) and the bottom graph showing the effects of a 3$\times$ reduction ($\gamma_{\rm amp,b} = 1.05$ \text{THz}). 10,000 shots are performed for each case.}
    \label{fig:3-site-amplitude-damping}
\end{figure}

{\em Amplitude Damping Effects}: Fig.~\ref{fig:3-site-amplitude-damping} depicts the population dynamics under different amplitude damping rates for the three chromophores, modeled using the Lindblad jump operator $\sigma^+$. The top panel compares a system-wide damping rate of $\gamma_{\rm amp, all}=3.15\times10^{12}$ Hz (defined in Table \ref{tab:params-final-H} against a non-dissipative reference evolution (dashed lines). 

These results indicate a substantial decrease in the chromophore excited state population, with only 21~\% and 4~\% of the initial population remaining at 0.5 and 1 ps, respectively. These values closely match the theoretical expectation: after $t/\tau$ Trotter steps, the undamped population follows 
\begin{equation}
    (1-p_{\rm amp,all})^{t/\tau} = \left(e^{-\gamma_{\rm amp,all}\tau}\right)^{t/\tau} = e^{-\gamma_{\rm amp,all}t},
\end{equation} 
also yielding 21~\% and 4~\% at 0.5 and 1 ps, respectively. 

Furthermore, adjusting individual chromophore damping rates (e.g.,  changing the local chemical environment of the chromophore) offers a potential mechanism for controlling energy transfer pathways. The middle and bottom panels of Fig.~\ref{fig:3-site-amplitude-damping} illustrate the effects of increasing and decreasing the damping rate of chromophore $B$ by a factor of three ($\gamma_{\rm amp,b} = 9.45\times10^{12}$ Hz and $\gamma_{\rm amp,b} = 1.05\times10^{12}$ Hz, respectively).  As expected, increasing (decreasing) the damping rate leads to a lower (higher) excited-state population for chromophore $B$. Notably, this tuning also temporarily enhances (supresses) the excited-state populations of chromophores $A$ and $C$, suggesting a transient redistribution of energy before ultimate dissipation. We hypothesize that a reduced damping rate on $B$ allows energy to accumulate and subsequently transfer to $A$ and $C$ before environmental dissipation dominates.

{\em Dephasing Effects}: Dephasing, the second dissipation mechanism under investigation, leads to quantum coherence loss without energy dissipation,~\cite{Nielsen_Chuang} causing the system to evolve toward a mixed state over time.

\begin{figure}[htbp]
    \centering
    \includegraphics[width=0.6 \textwidth]{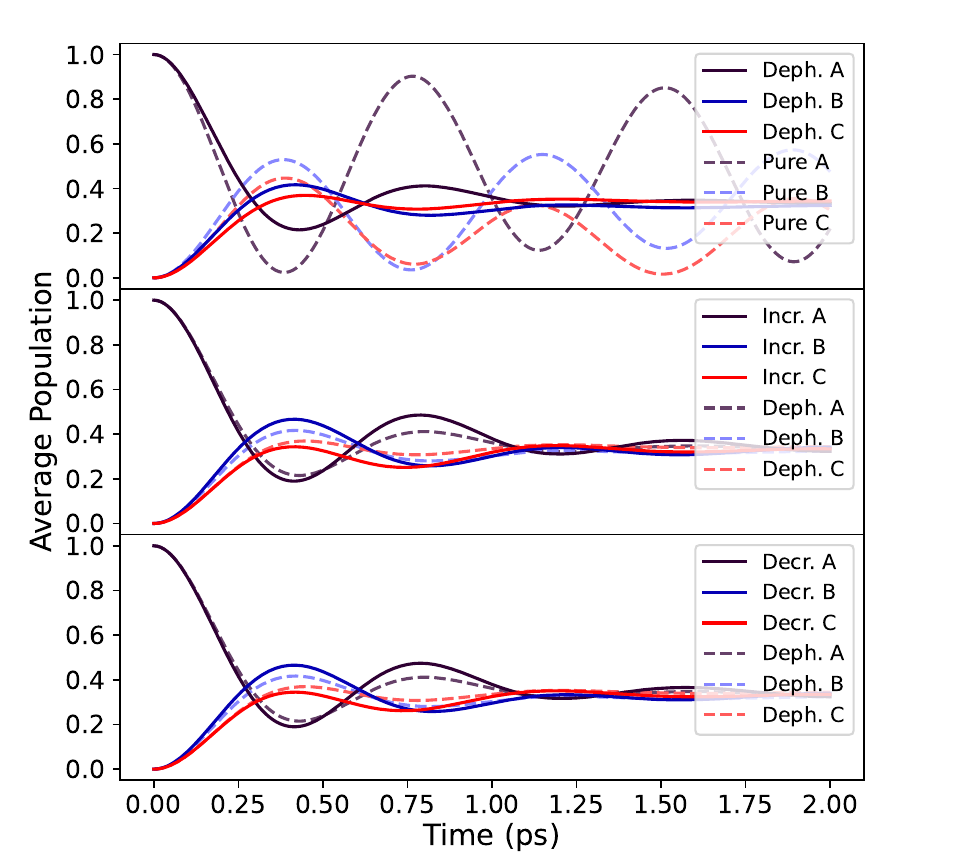}
    \caption{Population dynamics of the 3-site chromophore system under different dephasing rates. The top panel shows dephasing dissipation $\gamma_{\rm dep,all}= 0.9$ THz at 77~K, plotted against a non-dissipative system. The middle and bottom panels demonstrate the effects of tuning dephasing rate on chromophore $B$, with the middle panel showing a 3$\times$ increase ($\gamma_{\rm dep,b} = 2.7$ THz at 231~K) and the bottom panel showing a 3$\times$ reduction ($\gamma_{\rm dep,b} = 0.3$ THz at 25.6~K) with respect to a reference dephasing simulation (dashed lines in the middle and lower panels). 10,000 shots are performed for each case.}
    \label{fig:3-site-dephasing}
\end{figure}

At higher temperatures, dephasing rates increase, accelerating the relaxation of the system.~\cite{cqed-vibronic-sim} For instance, in the spin-boson model discussed in Sec.~\ref{sec:diss}, the dephasing rate is inversely proportional to the inverse temperature, $\beta=1/kT$ (Table~\ref{tab:para_lindblad_SBM}). For this analysis, we select a physically relevant dephasing rate of $\gamma_{\rm dep} = 9.0\times 10^{11}$~Hz, corresponding to an experimental system temperature of approximately $77$~K.~\cite{Pelzer2012, Panitchaya2010} Applying this rate on the 3-site chromophore system and comparing it to the dissipationless (top panel of Fig.~\ref{fig:3-site-dephasing}), we observe that the system decays as expected to a mixed state while maintaining the total excited-state chromophore population. 

To explore the effect of selective dephasing, we vary the dephasing rate of chromophore $B$ (middle and bottom panels of Fig.~\ref{fig:3-site-dephasing}). A higher dephasing rate accelerates relaxation while reducing the transient population of excited-state chromophore $B$, whereas a lower dephasing rate results in slower relaxation and higher transient excited-state populations. This behavior can be attributed to the nature of phase damping: since dephasing does not dissipate energy into the environment, the excited-state population redistributes across the chromophores as coherence is lost.

\begin{figure}[htbp]
    \centering
    \includegraphics[width=0.6\textwidth]{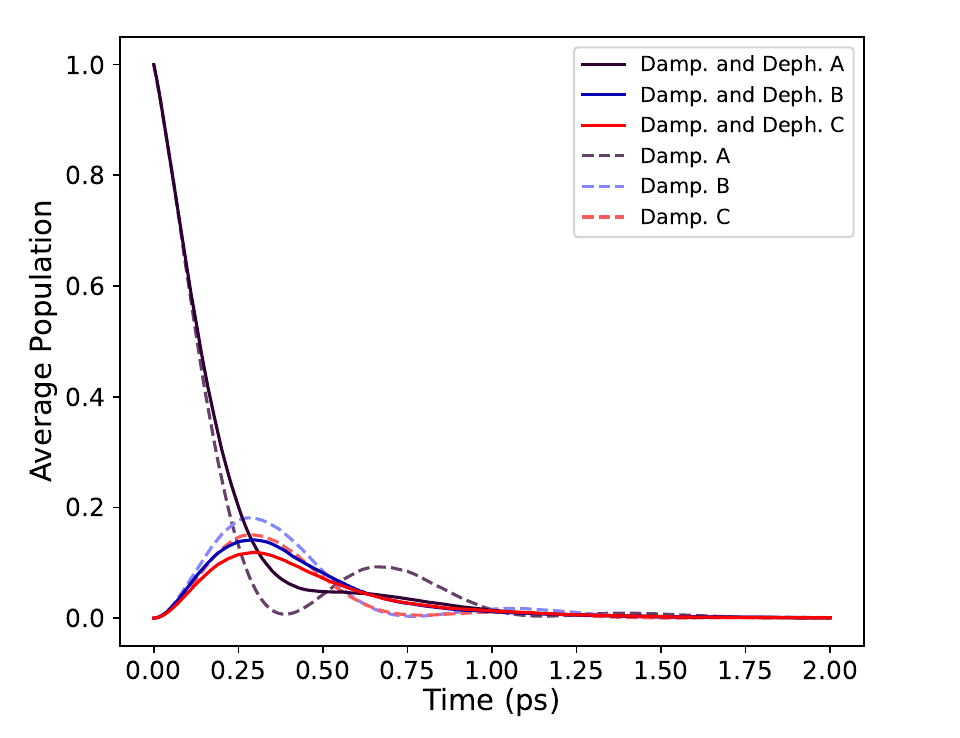}
    \caption{Population dynamics of the 3-site chromophore system under both dephasing and amplitude damping, compared to a system with amplitude damping only. The damping rates used are $\gamma_{\rm amp,all}= 3.15\times 10^{12}~\text{Hz}$ and $\gamma_{\rm dep,all} = 9.0\times 10^{11}~\text{Hz}$, as defined in Table~\ref{tab:params}. Each data point represents an average over 10,000 measurement shots.}
    \label{fig:3-site-dephased-and-damped}
\end{figure}

{\em Combined Amplitude Damping and Dephasing}: To achieve a more comprehensive and physically relevant simulation, we incorporate both amplitude damping and dephasing effects in the 3-chromophore system, as shown in Fig.~\ref{fig:3-site-dephased-and-damped}. 
Comparing the damped-dephased system with the damped-only case highlights the additional influence of environmental dephasing. The results indicate that the presence of both amplitude and phase damping suppresses most oscillations in the excited-state population, leading to a single peak for chromophores $B$ and $C$. This suggests that dephasing accelerates relaxation, reducing the coherence-driven oscillations observed in purely damped systems. 

\subsection{Noise Tolerance and Analysis}
\label{ssec:noise-analysis}
Current state-of-the-art quantum hardware is subject to three primary sources of error: gate infidelity, decoherence from thermal relaxation and dephasing, and state preparation and measurement (SPAM) errors.~\cite{Georgopoulos2021Noise} However, in the context of the 3-site chromophore system, the dominant hardware challenges arise from noisy controlled-NOT (CNOT) and conditional displacement (CD) operations, as indicated by Eqs.~\eqref{eq:gate_count}-\eqref{eq:alt_gate_count}. 

In Appendix \ref{asec:CNOT-noisy}, we simulate the population dynamics of both the dissipative and non-dissipative 3-chromophore systems in the presence of various CNOT infidelity levels. We then demonstrate that the dominant energy transfer pathway can still be determined if the infidelity is approximately no larger than $10^{-4}$. In Appendix \ref{asec:fidelity-of-cD-gate}, we show that the parameter regime describing vibronic couplings in Eq.~\eqref{eq:final-ham-3site} is compatible with hardware implementation of high-fidelity CD operations that does not notably affect the population dynamics of the 3-site chromophore system. Thus, our proposed framework for vibronic dynamics simulation is robust against hardware noise that can be achieved with near-term quantum devices.

\section{Conclusion and Outlook} \label{sec:conclusion}
We have introduced a general framework for simulating vibronic dynamics in chromophore arrays using programmable hybrid oscillator-qubit quantum hardware. Our approach incorporates energy dissipation into the simulation via engineered quantum channels, paving the way for co-designing gate-based quantum circuits applicable to both open and closed quantum systems. This work strengthens the link between high-level quantum algorithms and low-level hardware constraints, advancing towards a demonstration of quantum advantage in practical applications.

Starting with a trimer chromophore Hamiltonian inspired by photosynthetic antenna systems, we mapped the molecular Hamiltonian to the Hamiltonian of a cQED platform. We then generalized it to a one-dimensional multiple-site array. By encoding vibrational states in qumodes, we emulated the dynamics of bosonic modes involved in energy transfer, a computationally demanding task for quantum computers that rely solely on qubit platforms. 

For the hybrid CV-DV platflorm we demonstrated how amplitude damping and dephasing channels can be encoded to implement Lindblad dynamics. Based on this, we proposed a modular cQED hardware design and compiled the system Hamiltonian using a native instruction set architecture. Our numerical simulations confirmed that the vibronic population dynamics remained robust even in the presence of 0.01\% CNOT gate infidelity.

This work opens several avenues at the intersection of hardware-algorithm co-design and chemical physics. On the chemistry side, analogous quantum mappings could enable efficient simulations of reaction dynamics near conical intersections where the Born-Oppenheimer approximation breaks down.~\cite{yarkony1996diabolical,staab2022analytic} At the algorithmic level, while we focus on Trotterization and product formulas, investigating alternative approaches such as quantum signal processing and linear combination of unitaries will be necessary to determine the most efficient algorithms for specific hardware.

On the hardware front, novel platforms that enable scalable qumode implementations, such as multi-mode cavities,~\cite{Chakram2021Seamless} present promising opportunities  for vibronic simulations. Optimizing instruction set architectures for these platforms will be essential.~\cite{liu2024hybrid} While our results demonstrate viability under intermediate gate error rates, long-time simulations will require integrating error correction and mitigation strategies into the co-design process.~\cite{cai2021bosonic,Sivak_GKP_2022,albert_performance_2018} Finally, as chemical systems and quantum hardware grow increasingly complex, automated quantum compilers will become essential for scalable and efficient circuit design.\cite{chong2017programming, zhou2024bosehedral} We look forward to future developments along these directions.

\begin{acknowledgement}
We thank Benjamin Brock, Alec Eickbusch, and Di Luo for helpful discussions. Y.L. thanks Benjamin Brock and Alec Eickbusch for helping acquire the funding. Y.L. acknowledges the support from the Co-Design Center for Quantum Advantage (C$^2$QA), National Quantum Information Science Research Center via a Seed Funding Award under DE-SC0012704. D.D. and Y.L. are supported in part by the U.S. Department of Energy, Office of Science, Office of Advanced Scientific Computing Research (ASCR), under Award Number DE-SC0025384.  V.S.B. acknowledges the National Science Foundation Engines Development Award: Advancing Quantum Technologies (CT) under Award Number 2302908 and partial support from the National Science Foundation Center for Quantum Dynamics on Modular Quantum Devices (CQD-MQD) under Award Number 2124511. N.P.V. acknowledges support from the Lafayette College Dorflinger Summer Scholars and the Bergh Family Fellows Programs.
\end{acknowledgement}

\section*{Code and Data Availability}
All source and bench data, including runtimes, graphs, data processing, and analysis are available from the public GitHub repository:~\cite{cqed-vibronic-sim}
\begin{center}
    \url{https://github.com/yuanliu1/cqed-vibronic-simulation}.
\end{center}

\appendix

\section{Derivation of the cQED Effective Hamiltonian}\label{asec:H-derivation}
In this Appendix we provide a detailed derivation of the cQED effective Hamiltonian, given in Eq.~\eqref{eq:final-ham-3site}, corresponding to the model system Hamiltonian introduced by Eq.~\eqref{general-ham-5} with the parameters as provided in Table~\ref{tab:params}. 
\begin{table}[H]
    \centering
      \caption{Parameters for the three-chromophore antenna model, relevant to energy transfer in the photosynthetic process. Most values are adapted from the dimer chromophore analogue model in Ref.~\citenum{arsenault2021vibronic}. Parameters for chromophore $C$ are selected at the same order of magnitude with those in the dimer chromophore model within a physically relevant regime.}
    \label{tab:params}
    \begin{tabular}{|c|c|c|}
       \hline
       Parameters  & Values & Values (converted)\\
       \hline
       $\omega_{g, a}$ & 1650 $\mathrm{cm}^{-1}$ & $4.95\times 10^{13}$ Hz \\
       $\omega_{e, a}$ & 1545  $\mathrm{cm}^{-1}$ & $4.63\times 10^{13}$ Hz\\
       $\omega_{g, b}$ & 1660  $\mathrm{cm}^{-1}$ & $4.98\times 10^{13}$ Hz\\
       $\omega_{e, b}$ & 1540  $\mathrm{cm}^{-1}$ & $4.62\times 10^{13}$ Hz\\
       $\omega_{g, c}$ & 1640  $\mathrm{cm}^{-1}$ & $4.92\times 10^{13}$ Hz\\
       $\omega_{e, c}$ & 1550  $\mathrm{cm}^{-1}$ & $4.65\times 10^{13}$ Hz\\
       $\omega_{l}$ & 200  $\mathrm{cm}^{-1}$ & $6.00\times 10^{12}$ Hz\\
       $J_{AB, 0}$ & 100  $\mathrm{cm}^{-1}$ & $3.00\times 10^{12}$ Hz\\
       $J_{AC, 0}$ & 90  $\mathrm{cm}^{-1}$ & $2.70\times 10^{12}$ Hz\\
       $\eta_{AB}$ & -0.1 &\\
       $\eta_{AC}$ & 0.15 &\\
       $S_{a}$ & 0.005 &\\
       $S_{b}$ & 0.004 &\\
       $S_{c}$ & 0.006 &\\
       $S_{l}$ & 0 -- 0.1 (tunable) &\\
       $\gamma_{\rm amp,all}$ & 105  $\mathrm{cm}^{-1}$ & $3.15\times 10^{12}$ Hz\\
       $\gamma_{\rm dep,all}$ & 30  $\mathrm{cm}^{-1}$ & $9.00\times 10^{11}$ Hz\\
       \hline
    \end{tabular}
\end{table} 
We regroup Eq.~\eqref{general-ham-5} as follows: {\allowdisplaybreaks\begin{align}
    H&=  \underbrace{\left(\ket{G}\bra{G} + \ket{B}\bra{B} + \ket{C}\bra{C} \right) \otimes \hat{h}_A^{g} + \ket{A}\bra{A}\otimes \hat{h}^{e}_A}_{\mathcal{H}_a}  \nonumber \\
    &+ \underbrace{\left(\ket{G}\bra{G} + \ket{A}\bra{A} + \ket{C}\bra{C} \right) \otimes \hat{h}_B^{g} + \ket{B}\bra{B}\otimes \hat{h}^{e}_B}_{\mathcal{H}_b} \nonumber \\
    &+ \underbrace{\left(\ket{G}\bra{G} + \ket{A}\bra{A} + \ket{B}\bra{B} \right) \otimes \hat{h}_C^{g} + \ket{C}\bra{C}\otimes \hat{h}^{e}_C}_{\mathcal{H}_c}   \nonumber \\
    &+ \underbrace{J_{AB} \left( \ket{A}\bra{B} + h.c. \right) + J_{AC} \left( \ket{A}\bra{C} + h.c. \right)}_{\mathcal{J}}. \label{eq:asec-general-ham-rewrite}
\end{align}}
We reorder $\mathcal{H}_a$, defined by the first line of Eq.~\eqref{eq:asec-general-ham-rewrite}, as follows:{\allowdisplaybreaks\begin{align}
    \mathcal{H}_a &= I \otimes \hat{h}^g_A + \ket{A}\bra{A}\otimes \left(\hat{h}^e_A - \hat{h}^g_A\right) = I \otimes \hat{h}^g_A + \frac{1}{2}\left(I-\sigma^z_a\right)\otimes \left(\hat{h}^e_A - \hat{h}^g_A\right) \nn \\
    &= \frac{1}{2}  I \otimes \left(\hat{h}^g_A+\hat{h}^e_A\right) -\frac{1}{2}\sigma^z_a\otimes \left(\hat{h}^e_A - \hat{h}^g_A\right),
     \label{eq:ha}
\end{align}} 
where we used the closure relation $I=\ket{G}\bra{G}+\ket{A}\bra{A} +\ket{B}\bra{B} +\ket{C}\bra{C}$ in the single-excitation manifold.

Eqs.~\eqref{eq:general_hgA}-\eqref{eq:general_hgR} allow us to expand Eq.~\ref{eq:ha}, as follows:
{\allowdisplaybreaks\begin{align}
    \frac{\mathcal{H}_a}{\hbar} &= \underbrace{\frac{\omega_{g,a}+\omega_{e,a}}{2}}_{\omega_a}\left(a^\dagger a+\frac{1}{2}\right) + \omega_l \left(l^\dagger l+\frac{1}{2}\right) + \frac{\omega_{e,a}S_a +\omega_l S_l}{2}  \nn \\
    &- \underbrace{\sqrt{S_a}\omega_{e,a}}_{g_{a}} \frac{1}{2} \left(a^\dagger+a\right) - \underbrace{\sqrt{S_l}\omega_l}_{g_{cd,l}} \frac{1}{2}\left(l^\dagger+l\right) + \sqrt{S_a}\omega_{e,a}\frac{\sigma^z_a}{2} \left(a^\dagger+a\right) + \sqrt{S_l}\omega_l\frac{\sigma^z_a}{2}\left(l^\dagger+l\right) \nn \\
    &- \underbrace{(\omega_{e,a}-\omega_{g,a})}_{\chi_a} \frac{\sigma^z_a}{2}\left(a^\dagger a+\frac{1}{2}\right) - ( \omega_{e,a}S_a+\omega_l S_l)\frac{\sigma^z_a}{2} \label{eq:asec_Ha}
\end{align}}
We first omit the global phase terms in Eq.~\eqref{eq:asec_Ha} during the time evolution $\exp\left(-\frac{i}{\hbar} Ht\right)$. Then, we perform a (time-independent) displaced frame transformation associated with the unitary \begin{equation}
    U_a = {\mathrm D}_a^\dagger(\upsilon)=\exp\left(\upsilon^*a-\upsilon a^\dagger\right).
\end{equation} Effectively, this transformation displaces chromophore A's high-frequency vibrational mode in the phase-space coordinates alongside its ladder operators by $\upsilon$: \begin{equation}
    a \mapsto U_a aU_a^\dagger = a+\upsilon, \quad\quad
    a^\dagger \mapsto U_a a^\dagger U_a^\dagger = a^\dagger + \upsilon^*
\end{equation}
and modifies the Hamiltonian as \begin{equation}
    \mathcal{H}_a \mapsto \tilde{\mathcal{H}}_a =U_a\mathcal{H}_aU_a^\dagger + (i\hbar)\underbrace{(\partial_t U_a)}_{0}U_a^\dagger ={\mathrm D}_a^\dagger(\upsilon)\mathcal{H}_a{\mathrm D}_a(\upsilon) 
\end{equation}
That is, for real values of $\upsilon$, \begin{align}
    \frac{\tilde{\mathcal{H}}_a}{\hbar} &= \omega_a a^\dagger a + \omega_l l^\dagger l -\chi_a \frac{\sigma^z_a}{2}a^\dagger a  -\left(\chi_a|\upsilon|^2 + \frac{\chi_a}{2}-2g_a\upsilon+ \omega_{e,a}S_a+\omega_l S_l\right)\frac{\sigma^z_a}{2} \nn \\
    &+\left(g_a-\chi_a \upsilon\right)\frac{\sigma^z_a}{2}\left(a^\dagger+a\right) + g_{cd,l}\frac{\sigma^z_a}{2} \left(l^\dagger+l\right) + \left(\omega_a \upsilon - \frac{g_a}{2}\right)\left(a^\dagger+a\right) -\frac{g_{cd,l}}{2}\left(l^\dagger + l\right) \label{eq:asec-Ha-rot-a}
\end{align}
By choosing $\upsilon=\frac{g_a}{2\omega_a}$, we have (numerically) canceled the classical part of the oscillator $a$'s phase-space trajectory described by the term proportional to $(a^\dagger + a)$ in the first term of the last line of Eq.~\eqref{eq:asec-Ha-rot-a}. Similarly, we perform a second displaced frame transformation on the low-frequency vibrational mode $l$ of chromophore $A$, associated with \begin{equation}
    U_l = {\mathrm D}_l^\dagger\left(\frac{g_{cd,l}}{2\omega_l}\right),
\end{equation} 
we can also cancel the classical part of oscillator $l$'s phase-space trajectory, simplifying Eq.~\eqref{eq:asec-Ha-rot-a} to 
{\allowdisplaybreaks\begin{align}
    \frac{\tilde{\mathcal{H}}_{al}}{\hbar} &=  \omega_a a^\dagger a  + \omega_l l^\dagger l - \chi_a\frac{\sigma^z_a}{2} a^\dagger a + \underbrace{g_a\frac{\omega_{g,a}}{\omega_a}}_{g_{cd,a}}\frac{\sigma^z_a}{2} \left(a^\dagger+a\right) + g_{cd,l}\frac{\sigma^z_a}{2} \left(l^\dagger+l\right) \nn \\
    &- \underbrace{\left(\omega_{e,a}S_a+\omega_l S_l + \frac{\chi_a}{2} + \chi_a\frac{g_a^2}{4\omega_a^2} -\frac{g_a^2}{\omega_a}-\frac{g_{cd,l}^2}{\omega_l^2}\right)}_{\omega_{qa}} \frac{\sigma^z_a}{2}.\label{eq:asec_Ha_final}
\end{align}} Repeating the similar process for $\mathcal{H}_b$ and $\mathcal{H}_c$ yields the displaced-frame Hamiltonians 
{\allowdisplaybreaks
\begin{align}
    &\frac{\tilde{\mathcal{H}}_b}{\hbar} = \omega_b b^\dagger b - \chi_b \frac{\sigma_b^z}{2} b^\dagger b + \underbrace{g_b\frac{\omega_{g,b}}{\omega_b}}_{g_{cd,b}} \frac{\sigma^z_b}{2} \left(b^\dagger + b\right) - \underbrace{\left(\omega_{e,b}S_b + \frac{\chi_b}{2} + \chi_b \frac{g_b^2}{4\omega_b^2}-\frac{g_b^2}{\omega_b}\right)}_{\omega_{qb}} \frac{\sigma^z_b}{2},\label{eq:asec_Hb_final} \\
    &\frac{\tilde{\mathcal{H}}_c}{\hbar} = \omega_c c^\dagger c - \chi_c \frac{\sigma_c^z}{2} c^\dagger c + \underbrace{g_c \frac{\omega_{g,c}}{\omega_c}}_{g_{cd,c}}\frac{\sigma^z_c}{2} \left(c^\dagger + c\right) - \underbrace{\left(\omega_{e,c}S_c + \frac{\chi_c}{2} + \chi_c \frac{g_c^2}{4\omega_c^2}-\frac{g_c^2}{\omega_c} \right)}_{\omega_{qc}} \frac{\sigma^z_c}{2}. \label{eq:asec_Hc_final}
\end{align}}
Then, within the single-quanta excitation manifold for the three qubits, 
\begin{equation}
    \ket{A}\bra{R} + \ket{R}\bra{A} = \ket{e}_A\bra{g}_A \otimes \ket{g}_R\bra{e}_R +  \ket{g}_A\bra{e}_A \otimes \ket{e}_R\bra{g}_R = \sigma^+_A \sigma^-_R + \sigma^-_A \sigma^+_R,
\end{equation}
for $R = B, C$, from which the energy hopping terms in $\mathcal{J}$, combined with Eq.~\eqref{eq:general_ham_JAR}, are equivalent to \begin{equation}
    \frac{\mathcal{J}}{\hbar} = \sum_{R=B,C} \biggl( J_{AR,0}\left(\sigma_A^-\sigma_R^+ + \sigma_A^+\sigma_R^-\right) + J_{AR,0}\eta_{AR}\left(\sigma_A^-\sigma_R^+ + \sigma_A^+\sigma_R^-\right) \left(l^\dagger+l\right) \biggl) \label{eq:asec_J}
\end{equation}
Eqs.~\eqref{eq:asec_Ha_final}-\eqref{eq:asec_Hc_final} and Eq.~\eqref{eq:asec_J} have led us to the displaced full system Hamiltonian
{\allowdisplaybreaks
\begin{align}
    \frac{\tilde{H}}{\hbar} &= \frac{\tilde{\mathcal{H}}_{al}}{\hbar} + \frac{\tilde{\mathcal{H}}_b}{\hbar} + \frac{\tilde{\mathcal{H}}_c}{\hbar} +\frac{\mathcal{J}}{\hbar} \nn \\
    &= \omega_a a^\dagger a + \omega_b b^\dagger b + \omega_c c^\dagger c + \omega_l l^\dagger l - \omega_{qa} \frac{\sigma_a^z}{2} - \omega_{qb} \frac{\sigma_b^z}{2} - \omega_{qc} \frac{\sigma_c^z}{2} \nn \\
    &- \frac{\chi_a}{2} a^\dagger a \sigma_a^z - \frac{\chi_b}{2} b^\dagger b \sigma_b^z - 
    \frac{\chi_c}{2} c^\dagger c \sigma_c^z + g_{ab} \left(\sigma^-_A \sigma^+_B + \sigma^+_A \sigma^-_B\right) + g_{ac} \left(\sigma^-_A \sigma^+_C + \sigma^+_A \sigma^-_C\right) \nn \\
    &+ g_{cd,a} (a + a^\dagger)\frac{\sigma_a^z}{2} + g_{cd,b} (b + b^\dagger)\frac{\sigma_b^z}{2} + g_{cd,c} (c + c^\dagger)\frac{\sigma_c^z}{2} 
    + g_{cd,l} (l + l^\dagger)\frac{\sigma_a^z}{2} \nn  \\
    &+ g_{abl} \left(\sigma^-_A \sigma^+_B + \sigma^+_A \sigma^-_B\right) (l + l^\dagger) + g_{acl} \left(\sigma^-_A \sigma^+_C + \sigma^+_A \sigma^-_C\right) (l + l^\dagger)
\end{align}
}
We now transform this Hamiltonian into the first rotating frame where the qubits $a,b,c$ rotate at frequencies $\omega_{qa},\omega_{qb}$, and $\omega_{qc}$, respectively. This results in the detuning frequencies of $\Delta_r = 0$ for all qubits $r=a,b,c$ and effectively transforms \begin{equation}
    \sigma^\pm_R \mapsto \sigma^\pm_R e^{\pm i\omega_{qr} t}
\end{equation} for $R = A,B,C$. The Hamiltonian now has become 
{\allowdisplaybreaks\begin{align}
    \frac{\tilde{H}}{\hbar} &= \omega_a a^\dagger a + \omega_b b^\dagger b + \omega_c c^\dagger c + \omega_l l^\dagger l \nonumber -\frac{\chi_a}{2} a^\dagger a \sigma_a^z - \frac{\chi_b}{2} b^\dagger b \sigma_b^z - 
    \frac{\chi_c}{2} c^\dagger c \sigma_c^z \nonumber  \\
    &+ g_{cd,a} (a + a^\dagger)\frac{\sigma_a^z}{2} + g_{cd,b} (b + b^\dagger)\frac{\sigma_b^z}{2} + g_{cd,c} (c + c^\dagger)\frac{\sigma_c^z}{2} 
    + g_{cd,l} (l + l^\dagger)\frac{\sigma_a^z}{2} \nonumber  \\
    &+ g_{ab} \left(\sigma^-_A \sigma^+_Be^{-i\Delta_{ab}t} + \sigma^+_A \sigma^-_Be^{i\Delta_{ab}t}\right) + g_{ac} \left(\sigma^-_A \sigma^+_Ce^{-i\Delta_{ac}t} + \sigma^+_A \sigma^-_Ce^{i\Delta_{ac}t}\right) \nonumber \\
    &+ g_{abl} \left(\sigma^-_A \sigma^+_B e^{-i\Delta_{ab}t} + \sigma^+_A \sigma^-_B e^{i\Delta_{ab}t}\right) (l + l^\dagger) + g_{acl} \left(\sigma^-_A \sigma^+_Ce^{-i\Delta_{ac}t} + \sigma^+_A \sigma^-_C e^{i\Delta_{ac}t}\right) (l + l^\dagger)
\end{align}}
where $\Delta_{xy} = \omega_{qx} - \omega_{qy}$. We remark from this transformation that only the relative difference between qubit frequencies are relevant for the system dynamics at stake. 
With this in mind, we now make a second rotating frame transformation, to ``re-absorb" the time dynamics into a static Hamiltonian where we consider qubits $b$ and $c$ at relative frequencies $\Delta_{ab}$ and $\Delta_{ac}$, respectively. The composition of this rotating frame and the previous one is equivalent to a rotating frame transformation from the original system Hamiltonian $H$ with frequency $\omega_{qa}$ for all qubits. We then obtain the static Hamiltonian:
{\allowdisplaybreaks
\begin{align}
    \frac{\tilde{H}}{\hbar} &= \omega_a a^\dagger a + \omega_b b^\dagger b + \omega_c c^\dagger c + \omega_l l^\dagger l - \Delta_{ab} \frac{\sigma_b^z}{2} - \Delta_{ac} \frac{\sigma_c^z}{2} \nn \\
    & -\frac{\chi_a}{2} a^\dagger a \sigma_a^z - \frac{\chi_b}{2} b^\dagger b \sigma_b^z - 
    \frac{\chi_c}{2} c^\dagger c \sigma_c^z + g_{ab} \left(\sigma^-_A \sigma^+_B + \sigma^+_A \sigma^-_B\right) + g_{ac} \left(\sigma^-_A \sigma^+_C + \sigma^+_A \sigma^-_C\right) \nonumber \\
    &+ g_{cd,a} (a + a^\dagger)\frac{\sigma_a^z}{2} + g_{cd,b} (b + b^\dagger)\frac{\sigma_b^z}{2} + g_{cd,c} (c + c^\dagger)\frac{\sigma_c^z}{2} 
    + g_{cd,l} (l + l^\dagger)\frac{\sigma_a^z}{2} \nonumber \\
    &+ g_{abl} \left(\sigma^-_A \sigma^+_B + \sigma^+_A \sigma^-_B\right) (l + l^\dagger) + g_{acl} \left(\sigma^-_A \sigma^+_C + \sigma^+_A \sigma^-_C\right) (l + l^\dagger).
\end{align}
}
Finally, using the fact that 
\begin{align}
    \sigma^-_A \sigma^+_B = \frac{\sigma^x_a \sigma^x_b + \sigma^y_a \sigma^y_b}{2},
\end{align}
we arrive at the final rotating frame Hamiltonian given in Eq.~\eqref{eq:final-ham-3site}, with Table~\ref{tab:params-final-H} summarizing the experimental parameters of the system as described by these equations (frequencies are scaled for compatibility with the microwave domain).
\begin{table}[htbp]
    \centering
     \caption{Experimental parameters of the effective vibronic Hamiltonian in the cQED framework. Frequencies are in Hz; scaling assumes a base rate of $10^5$ Hz on cQED hardware, so the actual frequencies on experimental devices are obtained by dividing the values of the last column by $10^{5}$ to place them in the MHz microwave regime, which is implementable with state-of-the-art devices.~\cite{EickbuschECD}} Relevant values are calculated with $S_l=0.05$.
    \begin{tabular}{|c|P{40mm}|c|}
    \hline
    \textbf{cQED} & \textbf{Model} & \textbf{Exp. Value}\\ \hline
    $\omega_a$ & $\left(\omega_{g,a} + \omega_{e,a}\right)/2$ & $4.79\times 10^{13}$  \\
    $\omega_b$ & $\left(\omega_{g,b} + \omega_{e,b}\right)/2$ & $4.80\times 10^{13}$  \\
    $\omega_c$ &  $\left(\omega_{g,c} + \omega_{e,c}\right)/2$ & $4.79\times 10^{13}$  \\
    $\omega_l$ &  $\omega_l$ & $6.00\times 10^{12}$  \\
    $\chi_a$ & $\omega_{e,a}-\omega_{g,a}$ & $-3.20\times 10^{12}$ \\
    $\chi_b$ & $\omega_{e,b}-\omega_{g,b}$ & $-3.60\times 10^{12}$  \\
    $\chi_c$ & $\omega_{e,c}-\omega_{g,c}$ & $-2.70\times 10^{12}$  \\
    $\omega_{qa}$ & See Eq.~\eqref{eq:asec_Ha_final} & $-1.30\times 10^{12}$  \\
    $\omega_{qb}$ & See Eq.~\eqref{eq:asec_Hb_final} & $-1.80\times 10^{12}$ \\
    $\omega_{qc}$ & See Eq.~\eqref{eq:asec_Hc_final} & $-1.35\times 10^{12}$  \\
    $\Delta_{ab}$ & $\omega_{qa} - \omega_{qb}$ & $5.00\times 10^{11}$ \\
    $\Delta_{ac}$ & $\omega_{qa} - \omega_{qc}$ & $4.99\times 10^{10}$ \\
    $g_{cd,a}$ & $\sqrt{S_a}\omega_{e,a}\omega_{g,a}/\omega_a$ & $3.38\times 10^{12}$  \\
    $g_{cd,b}$ & $\sqrt{S_b}\omega_{e,b}\omega_{g,b}/\omega_b$ & $3.03\times 10^{12}$  \\
    $g_{cd,c}$ & $\sqrt{S_c}\omega_{e,c}\omega_{g,c}/\omega_c$ & $3.70\times 10^{12}$  \\
    $g_{cd,l}$ & $\sqrt{S_l}\omega_{l}$ & $1.34\times 10^{12}$  \\
    $g_{ab}$ & $J_{AB,0}$ & $3.00\times 10^{12}$  \\
    $g_{ac}$ & $J_{AC,0}$ & $2.70\times 10^{12}$  \\
    $g_{abl}$ & $J_{AB,0}\eta_{AB}$ & $-3.00\times 10^{11}$  \\
    $g_{acl}$ & $J_{AC,0}\eta_{AC}$ & $4.05\times 10^{11}$ \\
    $\gamma_{\rm amp,all}$ & $\gamma_{\rm amp,all}$ & $3.15\times 10^{12}$  \\
    $\gamma_{\rm dep,all}$ & $\gamma_{\rm dep,all}$ & $9.00\times 10^{11}$ \\
    \hline
\end{tabular}
    \label{tab:params-final-H}
\end{table}

\section{Engineering Dissipation via Channel Dilation} \label{asec:channel-derivation}

The discussion of Lindbladian dynamics in Sec.~\ref{sec:diss} sets the stage for constructing quantum channels, which we now detail within the framework of gate-based quantum hardware. Consider the amplitude damping channel characterized by a damping probability $p$. The corresponding Kraus operators are: 
\begin{equation} A_0 = \sqrt{p} \ket{0}\bra{1}, \quad A_1 = \ket{0}\bra{0} + \sqrt{1 - p} \ket{1}\bra{1}. \end{equation}
Here, $A_0$ represents the relaxation of the excited state $\ket{1}$ to the ground state $\ket{0}$ while $A_1$ accounts for the partial preservation of the excited state population on the ground state. To ensure the map is physically valid, the set of Kraus operators $\{A_k\}$ must statisfy the completely positive and trace-preserving (CPTP) condition: 
\begin{equation}
    \sum_k A^\dagger_k A_k = I,
\end{equation} where $I$ is the identity operator.
To derive an isometric extension of this channel, we define an isometry $U^N_{A\rightarrow BE}$ that maps the system $A$ to a larger Hilbert space $BE$ comprising the system $B$ and the environment $E$:
\begin{equation} 
    U^N_{A\rightarrow BE} = \left(\sqrt{1-p}\ket{1}\bra{1} + \ket{0}\bra{0}\right)\otimes\ket{0}_E + \left(\sqrt{p}\ket{0}\bra{1}\right)\otimes\ket{1}_E .\label{eq:isometric_ext_amplitude_damping}
\end{equation}
This isometry (rectangular matrix) can be embedded into a unitary operation $V_{AE}$ (square matrix) on the combined system-environment space by extending the isometric matrix to a full unitary matrix through the addition of (two more) orthogonal columns:
\begin{equation}
    V_{AE} =
    \left[\begin{array}{cccc}
        1 & 0 & 0 & 0\\
        0 & 0 & \sqrt{p}  & \sqrt{1-p}\\
        0 & 0 & \sqrt{1-p}  & -\sqrt{p}\\
        0 & 1 & 0 & 0
    \end{array}\right] = \left[\begin{array}{cccc}
        1 & 0 & 0 & 0\\
        0 & 0 & \sin{(\theta/2)} & \cos{(\theta/2)}\\
        0 & 0 & \cos{(\theta/2)}  & -\sin{(\theta/2)}\\
        0 & 1 & 0 & 0
    \end{array}\right], \label{eq:full_unitary_amplitude_damping}
\end{equation}
where the second equality holds for $p = \sin^2(\theta/2)$. This parametrization facilitates an efficient gate-based realization of the amplitude damping process. 
The corresponding quantum circuit implementation is depicted in Fig. \ref{fig:noise_circuit_SBM} (b), where the system qubit $\ket{\phi}$ interacts with an ancilla qubit initialized in the ground state $\ket{0}$, representing the environment.

A dephasing channel can be constructed analogously, defined by the map 
\begin{equation}
\rho\rightarrow(1-p)\rho + p\sigma^z\rho\sigma^z,
\end{equation}
where the phase flips with probability $p$.~\cite{Nielsen_Chuang} From this definition, we derive the CPTP set of Kraus operators: 
\begin{equation}
K_0 = \sqrt{p} \sigma^z, \hspace{.3cm} K_1 = \sqrt{1-p}I,
\end{equation}
corresponding to the isometric extension:
\begin{equation}
U^\mathcal{N}_{A\rightarrow BE}=\sqrt{1-p}\ket{\psi_A}\otimes\ket{0}_E + \sigma^z\sqrt{p}\ket{\psi_A}\otimes\ket{1}_E
\end{equation}
which can be extended to the full unitary representation:
{\allowdisplaybreaks\begin{equation}
    V_{AE} = \left[\begin{array}{cccc}
        \sqrt{1-p} & \sqrt{p} & 0 & 0\\
        \sqrt{p} & -\sqrt{1-p} & 0  & 0\\
        0 & 0 & \sqrt{1-p}  & \sqrt{p}\\
        0 & 0 & -\sqrt{p} & \sqrt{1-p}
    \end{array}\right] = \left[\begin{array}{cccc}
        \cos{(\frac{\theta}{2})} & \sin{(\frac{\theta}{2})} & 0 & 0\\
        \sin{(\frac{\theta}{2})} & -\cos{(\frac{\theta}{2})} & 0 & 0\\
        0 & 0 & \cos{(\frac{\theta}{2})}  & \sin{(\frac{\theta}{2})}\\
        0 & 0 & -\sin{(\frac{\theta}{2})} & \cos{(\frac{\theta}{2})}
    \end{array}\right] \label{eq:full_unitary_dephasing}
\end{equation}}
as we have introduced the substitution $p = \sin^2(\theta/2)$. 

The corresponding quantum circuit, denoted as $\mathcal{E}_{\rm dep}$, is shown in Fig.~\ref{fig:noise_circuit_SBM} (c). Here, the rotation $\mathrm{R}_y(\theta)$ can be decomposed as $\mathrm{R}_y(-\theta)= \sigma^z \mathrm{R}_y(-\theta)$ when acting on the environment in the ground state $\ket{0}$, since \begin{equation}
    \mathrm{R}_y(\theta) \ket{0} = \sigma^z \mathrm{R}_y(-\theta)\ket{0}.
\end{equation}

\section{Compiling Quantum Circuits per Trotter Step} \label{asec:compiled_circuit_trotter}
In this Appendix we provide the full compilation to simulate each Trotter step $\tau$ for all the terms in Eq.~\eqref{total-ham-1d}, except those that describe dispersive vibronic coupling which are already covered in Sec.~\ref{sssec:compiled_circuit_trotter}.

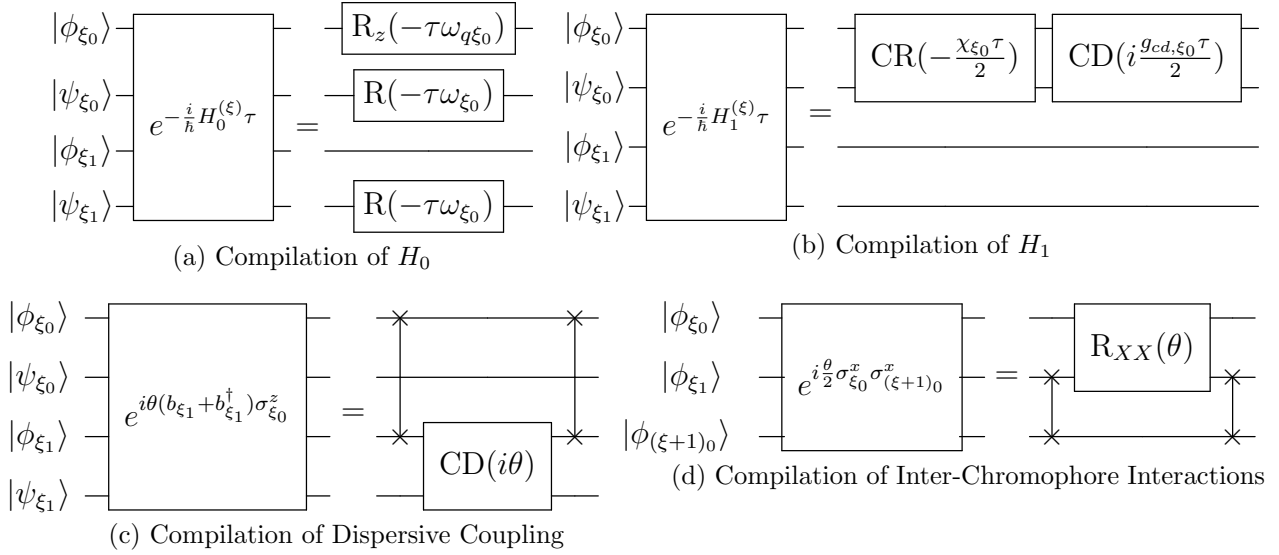
\begin{figure}[htbp]
    \centering
    \subfloat[Compilation of $H_0$]{
        \mbox{
            \Qcircuit @C=0.55em @R=.5em {
            \ket{\phi_{\xi_0}} &&& \multigate{3}{e^{-\frac{i}{\hbar}H_0^{(\xi)}\tau}} & \qw & 
                & \quad & \gate{\mathrm{R}_z(-\tau\omega_{q\xi_0})} & \qw \\
            \ket{\psi_{\xi_0}} &&& \oghost{e^{-\frac{i}{\hbar}H_0^{(\xi)}\tau}} & \ow & \raisebox{-2.25em}{=}
                & \quad & \gate{\mathrm{R}(-\tau\omega_{\xi_0})} & \ow \\
            \ket{\phi_{\xi_1}} &&& \ghost{e^{-\frac{i}{\hbar}H_0^{(\xi)}\tau}} & \qw & 
                & \quad & \qw & \qw \\
            \ket{\psi_{\xi_1}} &&& \oghost{e^{-\frac{i}{\hbar}H_0^{(\xi)}\tau}} & \ow & 
                & \quad & \gate{\mathrm{R}(-\tau\omega_{\xi_0})} & \ow
            }
        }
    }
    \quad
    \subfloat[Compilation of $H_1$]{
        \mbox{
            \Qcircuit @C=0.55em @R=1em {
            \ket{\phi_{\xi_0}} &&& \multigate{3}{e^{-\frac{i}{\hbar}H_1^{(\xi)}\tau}} & \qw & 
                & \quad & \multigate{1}{\mathrm{CR}(-\frac{\chi_{\xi_0}\tau}{2})} & \multigate{1}{\mathrm{CD}(i\frac{g_{cd,\xi_0}\tau}{2})} & \qw \\
            \ket{\psi_{\xi_0}} &&& \oghost{e^{-\frac{i}{\hbar}H_1^{(\xi)}\tau}} & \ow & \raisebox{-1.85em}{=}
                & \quad & \oghost{\mathrm{CR}(-\frac{\chi_{\xi_0}\tau}{2})} & \oghost{\mathrm{CD}(i\frac{g_{cd,\xi_0}\tau}{2})} & \ow \\
            \ket{\phi_{\xi_1}} &&& \ghost{e^{-\frac{i}{\hbar}H_1^{(\xi)}\tau}} & \qw & 
                & \quad & \qw & \qw & \qw \\
            \ket{\psi_{\xi_1}} &&& \oghost{e^{-\frac{i}{\hbar}H_1^{(\xi)}\tau}} & \ow & 
                & \quad & \ow & \ow & \ow
            }
        }
    } \\
    \subfloat[Compilation of Dispersive Coupling]{
        \mbox{
            \Qcircuit @C=.75em @R=1em {
            \ket{\phi_{\xi_0}} \quad \quad \quad & \multigate{3}{e^{i\theta (b_{\xi_1} + b^\dagger_{\xi_1})\sigma^z_{\xi_0}}} & \qw & 
                & \quad & \qswap & \qw & \qswap & \qw \\
            \ket{\psi_{\xi_0}} \quad \quad \quad & \oghost{e^{i\theta (b_{\xi_1} + b^\dagger_{\xi_1})\sigma^z_{\xi_0}}} & \ow & \raisebox{-2.5em}{=} && \qwx \ow & \ow & \qwx \ow & \ow\\
            \ket{\phi_{\xi_1}} \quad \quad \quad & \ghost{e^{i\theta (b_{\xi_1} + b^\dagger_{\xi_1})\sigma^z_{\xi_0}}} & \qw & 
                & \quad &\qswap \qwx & \multigate{1}{\text{CD}(i\theta)} & \qswap \qwx & \qw \\
            \ket{\psi_{\xi_1}} \quad \quad \quad & \oghost{e^{i\theta (b_{\xi_1} + b^\dagger_{\xi_1})\sigma^z_{\xi_0}}} & \ow & 
                & \quad & \ow & \oghost{\text{CD}(i\theta)} & \ow & \ow
            }
        }
    }
    \quad \quad
    \subfloat[Compilation of Inter-Chromophore Interactions]{
        \mbox{
            \Qcircuit @C=0.72em @R=1em {
            \ket{\phi_{\xi_0}} &&&& \multigate{2}{e^{i\frac{\theta}{2} \sigma^x_{\xi_0}\sigma^x_{(\xi+1)_0}}} & \qw &  
                & \quad & \qw & \multigate{1}{\mathrm{R}_{XX}(\theta)} & \qw & \qw \\
            \ket{\phi_{\xi_1}} &&&& \ghost{e^{i\theta \sigma^+_{\xi_0}\sigma^-_{(\xi+1)_0}}} & \qw & = 
                & \quad & \qswap & \ghost{\mathrm{R}_{XX}(\theta)} & \qswap & \qw \\
            \ket{\phi_{(\xi+1)_0}} \quad &&&& \ghost{e^{i\theta \sigma^+_{\xi_0}\sigma^-_{(\xi+1)_0}}} & \qw & 
                & \quad & \qswap \qwx & \qw & \qswap \qwx & \qw
            }
        }
    }
    \caption{(a) Compiling simulation of the term $H_0^{(\xi)}$ for the $\xi^\text{th}$ chromophore. (b) Compiling simulation of the term $H_1^{(\xi)}$ for the $\xi^\text{th}$ chromophore. (c) Compiling simulation of dispersive intra-chromophore interactions between the high-frequency electronic state \( \ket{\phi_{\xi_0}} \) and the low-frequency vibrational mode \( \ket{\psi_{\xi_1}} \) within the \( \xi^\text{th} \) chromophore. The interaction \( e^{i\theta (b_{\xi_1} + b^\dagger_{\xi_1}) \sigma^z_{\xi_0}} \) is decomposed into a conditional displacement gate \( \mathrm{CD}(i\theta) \) on the low-frequency mode, conjugated by SWAP operations that exchange the states of the high- and low-frequency transmons to facilitate the interaction. (d) Compiling simulation of inter-chromophore $\sigma^x\sigma^x$ interactions between high-frequency electronic states $\ket{\phi_{\xi_0}}$ and $\ket{\phi_{(\xi+1)_0}}$. Vibrational states encoded in qumodes $\ket{\psi_{\xi_0}}$ and $\ket{\psi_{(\xi+1)_0}}$ are omitted for brevity. The $\sigma^y\sigma^y$ interactions follow a similar structure, replacing the $\mathrm{R_{XX}}$ operation with $\mathrm{R_{YY}}$.}
    \label{fig:compile_H}
\end{figure}

{\em Compiling $H_0^{(\xi)}$}: The compilation of $H_0^{(\xi)}$ (Eq.~\eqref{eq:final-ham-msite-h0}) is straightforward, involving only single-qubit and single-qumode gates. The time-evolution of the terms involving the bosonic number operators $\omega_{\xi_0} b^\dagger_{\xi_0} b_{\xi_0}$ and $\omega_{\xi_1} b^\dagger_{\xi_1} b_{\xi_1}$ is implemented via phase-space rotation operations on the high- and low-frequency modes, respectively. The qubit term, $-\frac{\omega_{q\xi_0}}{2} \sigma^z_{\xi_0}$, corresponds to a Pauli-Z rotation applied on the high-frequency transmon qubit. The combined time-evolution operator is: 
\begin{align}
    e^{-\frac{i}{\hbar}H_0^{(\xi)}\tau} &\approx 
         e^{\frac{i}{\hbar}\frac{\omega_{q\xi_0}}{2} \sigma^z_{\xi_0}} \otimes e^{-\frac{i}{\hbar}\omega_{\xi_0} b^\dagger_{\xi_0} b_{\xi_0}} \otimes \mathbb{I}_{\xi_1} \otimes e^{-\frac{i}{\hbar}\omega_{\xi_1} b^\dagger_{\xi_1} b_{\xi_1}}, \nn \\
    &=  {\rm R_{z,\xi_0}\left(-\tau\omega_{q\xi_0}\right)} \otimes {\rm R_{\xi_0}\left(-\tau\omega_{\xi_0}\right)} \otimes \mathbb{I}_{\xi_1} \otimes {\rm R_{\xi_1}\left(-\tau\omega_{\xi_1}\right)}, 
\end{align} 
where $\mathbb{I}_{\xi_1}$ denotes the identity operation on the low-frequency transmons. Figure~\ref{fig:compile_H} (a) shows the corresponding quantum circuit for each chromophore $\xi$ evolving under $H_0$.

{\em Compiling $H_1^{(\xi)}$}: The term $H_1^{(\xi)}$ (Eq.~\eqref{eq:final-ham-msite-h1}) describes the vibronic interactions within within the high-frequency mode of each chromophore, corresponding to dispersive couplings between states $\ket{\phi_{\xi_0}}$ and $\ket{\psi_{\xi_0}}$. In the circuit topology (Fig.~\ref{fig:modular-cQED}), these states have direct connectivity, allowing efficient gate compilation. 

The term $-\frac{\chi_{\xi_0}}{2} b^\dagger_{\xi_0} b_{\xi_0} \sigma^z_{\xi_0}$ is implemented as a CR gate. The interaction $\frac{g_{cd, \xi_0}}{2} (b_{\xi_0} + b^\dagger_{\xi_0}) \sigma^z_{\xi_0}$ is then compiled as a CD operation. The approximate time-evolution operator, justified via the Trotter-Suzuki decomposition for small $\tau$, is: \begin{align}
    e^{-\frac{i}{\hbar}H_1^{(\xi)}\tau} &\approx \left(e^{\frac{i}{\hbar}\frac{g_{cd, \xi_0}\tau}{2} (b_{\xi_0} + b^\dagger_{\xi_0}) \sigma^z_{\xi_0}} e^{-\frac{i}{\hbar}\frac{\chi_{\xi_0}\tau}{2} b^\dagger_{\xi_0} b_{\xi_0} \sigma^z_{\xi_0}}\right) \otimes \mathbb{I}_{\xi_1}, \nn \\
    &= \left({\rm CD}_{\xi_0}\left(i\frac{g_{cd,\xi_0}\tau}{2}\right) {\rm CR}_{\xi_0}\left(-\frac{\chi_{\xi_0}\tau}{2}\right)\right) \otimes \mathbb{I}_{\xi_1}.
\end{align} 
Here, $\mathbb{I}_{\xi_1}$ indicates that no operation is performed on the low-frequency cavity and transmon. Fig.~\ref{fig:compile_H} (b) shows the corresponding circuit for each chromophore $\xi$ evolved with $H_1$.

{\em Compiling $H_2^{(\xi)}$}: The term $H_2^{(\xi)}$ (Eq.~\eqref{eq:final-ham-msite-h2}) describes vibronic transitions between adjacent chromophores, involving both intra- and inter-site couplings. In the cQED framework (Fig.~\ref{fig:modular-cQED}), these interactions are mediated via SNAIL couplers, which support only nearest-neighbor couplings. We consider the dispersive intra-chromophore coupling term
\begin{align} \frac{g_{cd, \xi_1}}{2} (b_{\xi_1} + b^\dagger_{\xi_1}) \sigma^z_{\xi_0}, \end{align}
which couples the high-frequency qubit $\ket{\phi_{\xi_0}}$ and the low-frequency vibrational mode $\ket{\psi_{\xi_1}}$ of the \( \xi^\text{th} \) chromophore. This interaction is implemented using a conditional displacement on the low-frequency mode, sandwiched between the SWAP operations that exchange the states of the high- and low-frequency transmons:\begin{align}
    e^{i\theta (b_{\xi_1} + b^\dagger_{\xi_1})\sigma^z_{\xi_0} } &= \swap_{\xi_0\xi_1} \cdot e^{i\theta\sigma^z_{\xi_1}\otimes\left(b_{\xi_1} + b^\dagger_{\xi_1}\right) } \cdot \swap_{\xi_0\xi_1}, \nonumber \\
    &= \swap_{\xi_0\xi_1} \cdot \mathrm{CD}_{\xi_1}(i\theta) \cdot \swap_{\xi_0\xi_1},
\end{align}
where $\theta = -\frac{g_{cd, \xi_1}\tau}{2}$. The corresponding quantum circuit is shown in Fig.~\ref{fig:compile_H} (c). 

{\em High-Frequency Electronic Coupling}: We now focus on the term, 
\begin{align}
\frac{g_{\xi_0, (\xi+1)_0}}{2} (\sigma^+_{\xi_0} \sigma^-_{(\xi+1)_0}),
\end{align}
which describes the high-frequency electronic coupling. Deriving how its time evolution can be simulated using an $XX$-rotation followed by another $YY$-rotation, both parametrized by 
\begin{align}
\theta = \frac{g_{\xi_0, (\xi+1)_0}\tau}{2},
\end{align}
is provided in Appendix~\ref{asec:H-derivation}. However, since the two high-frequency electronic states (qubits) are separated by a low-frequency mode, a pair of conjugate SWAP gates is required:
{\allowdisplaybreaks\begin{align}
    e^{i\theta\sigma^+_{\xi_0}\sigma^-_{(\xi+1)_0}} &= \enspace \swap_{\xi_1(\xi+1)_0} e^{i\frac{\theta}{2}\left(\sigma^x_{\xi_0}\sigma^x_{\xi_1}+\sigma^y_{\xi_0}\sigma^y_{\xi_1}\right)} \swap_{\xi_1(\xi+1)_0} \nn \\
    &\approx \enspace \swap_{\xi_1(\xi+1)_0} \mathrm{R}_{X_{\xi_0}X_{\xi_1}}(\theta) \mathrm{R}_{Y_{\xi_0}Y_{\xi_1}}(\theta)\swap_{\xi_1(\xi+1)_0} \nn \\
    &= \left[\swap_{\xi_1(\xi+1)_0} \mathrm{R}_{X_{\xi_0}X_{\xi_1}}(\theta) \swap_{\xi_1(\xi+1)_0}\right] \left[\swap_{\xi_1(\xi+1)_0}\mathrm{R}_{Y_{\xi_0}Y_{\xi_1}}(\theta)\swap_{\xi_1(\xi+1)_0}\right] \label{eq:compile_H2sigma+sigma-}
\end{align}}
where the approximation is justified via Trotterization for small $\theta$. Hardware constraints prevent simultaneous implementation of the $\mathrm{R_{XX}}$ and $\mathrm{R_{YY}}$ operations, requiring the separation into $\tilde{H}_{2,XX}/\hbar$ (Eq.~\eqref{eq:final-ham-3site-h2xx}, compiled in Fig.~\ref{fig:compile_H} (d)), and $\tilde{H}_{2,YY}/\hbar$ (Eq.~\eqref{eq:final-ham-3site-h2yy}) for the 3-site chromophore model. Finally, The term 
\begin{align}
\frac{g_{\xi_0, (\xi-1)_0}}{2} (\sigma^+_{\xi_0} \sigma^-_{(\xi-1)_0}),
\end{align}
is compiled analogously by decrementing $\xi$ in Eq.~\eqref{eq:compile_H2sigma+sigma-}.

\section{Simulation Performance and Convergence} \label{asec:Performance-Convergence}
Several parameters influence the performance of Trotter-based simulations on a cQED device, including the Trotter step size, Fock truncation level, and shot count. In this section, we systematically vary these parameters to assess their effects on the simulation accuracy and performance, ultimately determining optimal parameters.

{\em Accuracy Assessment}: To quantify accuracy, we compute the Root Mean Square Error (RMSE) for each parameter set by comparing five independent simulations against another set of five simulations, forming a fully connected bipartite graph with a total of 25 comparison points. These RMSE values are averaged and normalized against the average RMSE of the (intuitively) most accurate parameter in each class, yielding the normalized RMSE values presented in Table \ref{tab:simulation_parameters_nrmse}. The reference dataset is chosen as a median of all the comparison points, with a Trotter step size $\tau=10$~fs, 10,000 measurement shots, and a Fock truncation of 8 levels. 

Additionally, to provide a baseline for general simulation errors, we compute the normalized internal average RMSE (visualized as a fully connected graph of comparison points) for simulations using the most accurate parameter in each category.

{\em Parameter Scaling and Optimization}: We benchmark Trotter step sizes ranging from 5 to 40 fs (corresponding to 200 to 25 steps per ps). Since the RMSE variance is significantly influenced by $\tau$, minimizing the step size is desirable. Notably, the average normalized RMSE nearly halves when reducing $\tau$ from 40 fs to 20 fs, whereas the improvement from 20 fs to 10 fs is less pronounced but still substantial. While larger step sizes can provide a qualitative understanding of the system dynamics, a smaller $\tau$ should be chosen whenever it is computationally feasible. In the presence of hardware intrinsic noise, an optimal trade-off between Trotter error and hardware-induced errors should be considered in future experimental implementations.
\begin{table}[H]
    \centering
        \caption{Normalized RMSE for various simulation parameter sets. The first column presents results for varying Trotter step sizes, followed by shot counts and Fock truncation levels. All simulations include environmentally induced dissipation, as in Fig.~\ref{fig:3-site-dephased-and-damped} and are performed on the Lafayette College High Performance Cluster. We use $^*$ to denote the normalized RMSE calculated amongst itself, which provides a baseline.}
    \begin{tabular}{|p{4cm}||p{2.5cm}|p{2.5cm}|p{2.5cm}|}
         \hline
         Comparison & Chromo. $A$ & Chromo. $B$ & Chromo. $C$ \\
         \hline
         5 - 5 fs$^*$ & $1.8\%$ & $2.6\%$ & $3.7\%$ \\
         5 - 10 fs & $2.2\%$ & $3.6\%$ & $4.6\%$ \\
         5 - 20 fs & $2.4\%$ & $5.5\%$ & $6.9\%$ \\
         5 - 40 fs & $6.0\%$ & $11\%$ & $13\%$ \\
         \hline
         20,000 - 20,000 shots$^*$ & $1.4\%$ & $3.7\%$ & $3.2\%$ \\
         20,000 - 10,000 shots & $1.8\%$ & $3.8\%$ & $3.5\%$ \\
         20,000 - 5,000 shots & $3.1\%$ & $5.0\%$ & $4.5\%$ \\
         20,000 - 2,500 shots & $3.0\%$ & $7.2\%$ & $6.9\%$ \\
         \hline
         16 - 16 Fock levels$^*$ & $1.5\%$ & $2.6\%$ & $4.5\%$ \\
         16 - 8 Fock levels & $2.1\%$ & $3.4\%$ & $4.1\%$ \\
         16 - 4 Fock levels & $1.7\%$ & $4.0\%$ & $4.4\%$ \\
         16 - 2 Fock levels & $1.9\%$ & $4.3\%$ & $4.7\%$ \\
         \hline
    \end{tabular}
    \label{tab:simulation_parameters_nrmse}
\end{table}

{\em Measurement Shots}: Shot count influences simulation variance, as more measurements reduce statistical fluctuations. The error reduction trend is noticeable, though less significant than that observed with Trotter step size refinement. We observe that computational runtime scales linearly with shot count, yet multiple simulations can be averaged to achieve equivalent effects, i.e., optimizing shot count is not as critical.

{\em Fock Truncation Level}: Intuitively, Fock truncation level directly impacts the simulation accuracy, as a low truncation level can exclude essential aspects of the system dynamics while a high truncation level is computationally expensive. We observe from Table~\ref{tab:simulation_parameters_nrmse} that, with the exception of chromophore $B$, lower Fock truncation levels do not significantly impact the normalized RMSE specifically in our 3-chromophore dissipative system.

\section{CNOT-noisy Numerical Simulations} \label{asec:CNOT-noisy}
In this Appendix we explain how noisy CNOT operations are simulated and demonstrate that infidelities no larger than $10^{-4}$ suffice to determine the dominant energy transfer pathway in the 3-site chromophore system.
\begin{figure}[htbp]
    \centering
    \subfloat[No Dissipative Noise]{\includegraphics[width=0.5\textwidth]{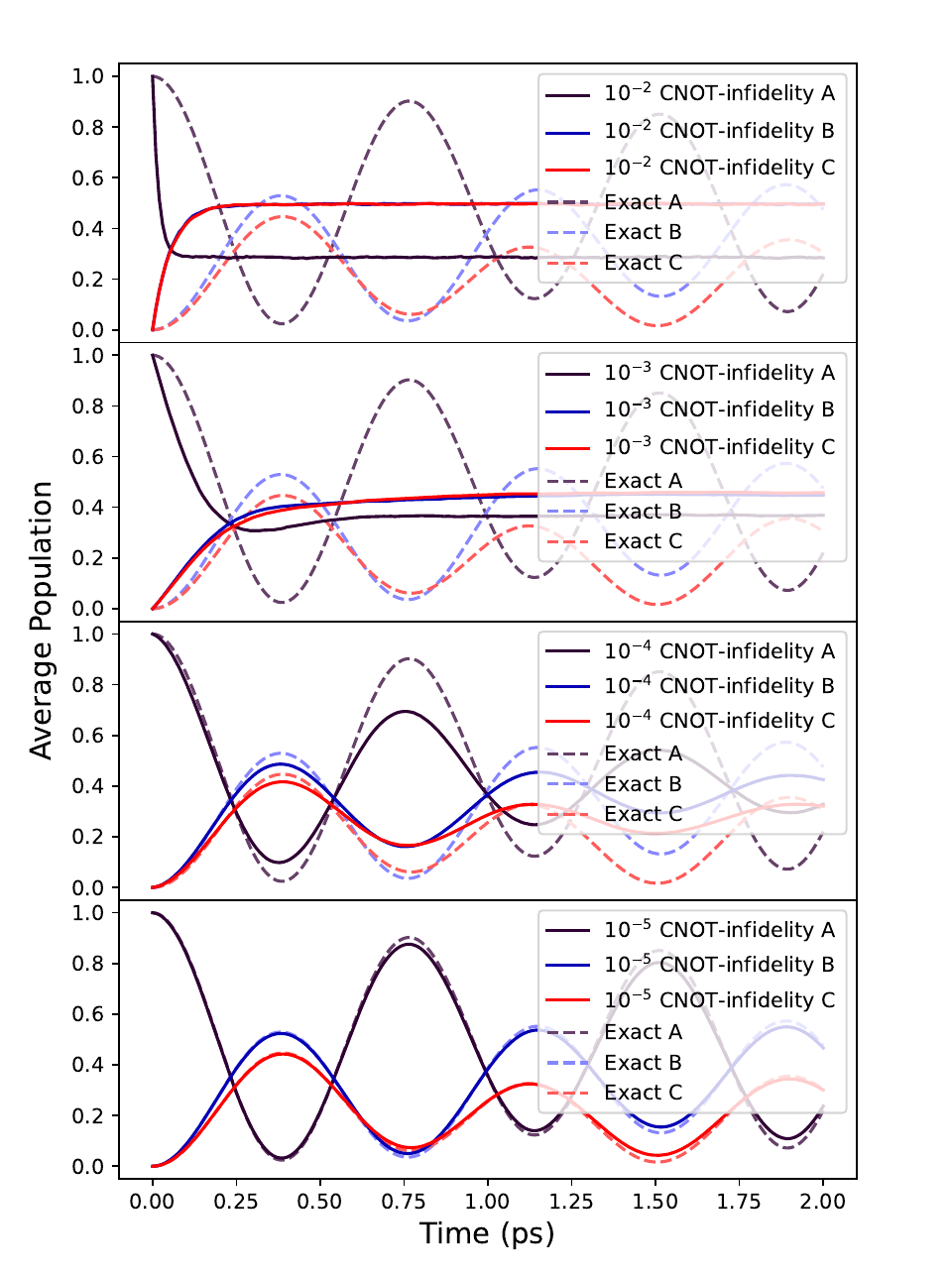}}
    \subfloat[Disspative Noise]{\includegraphics[width=0.5\textwidth]{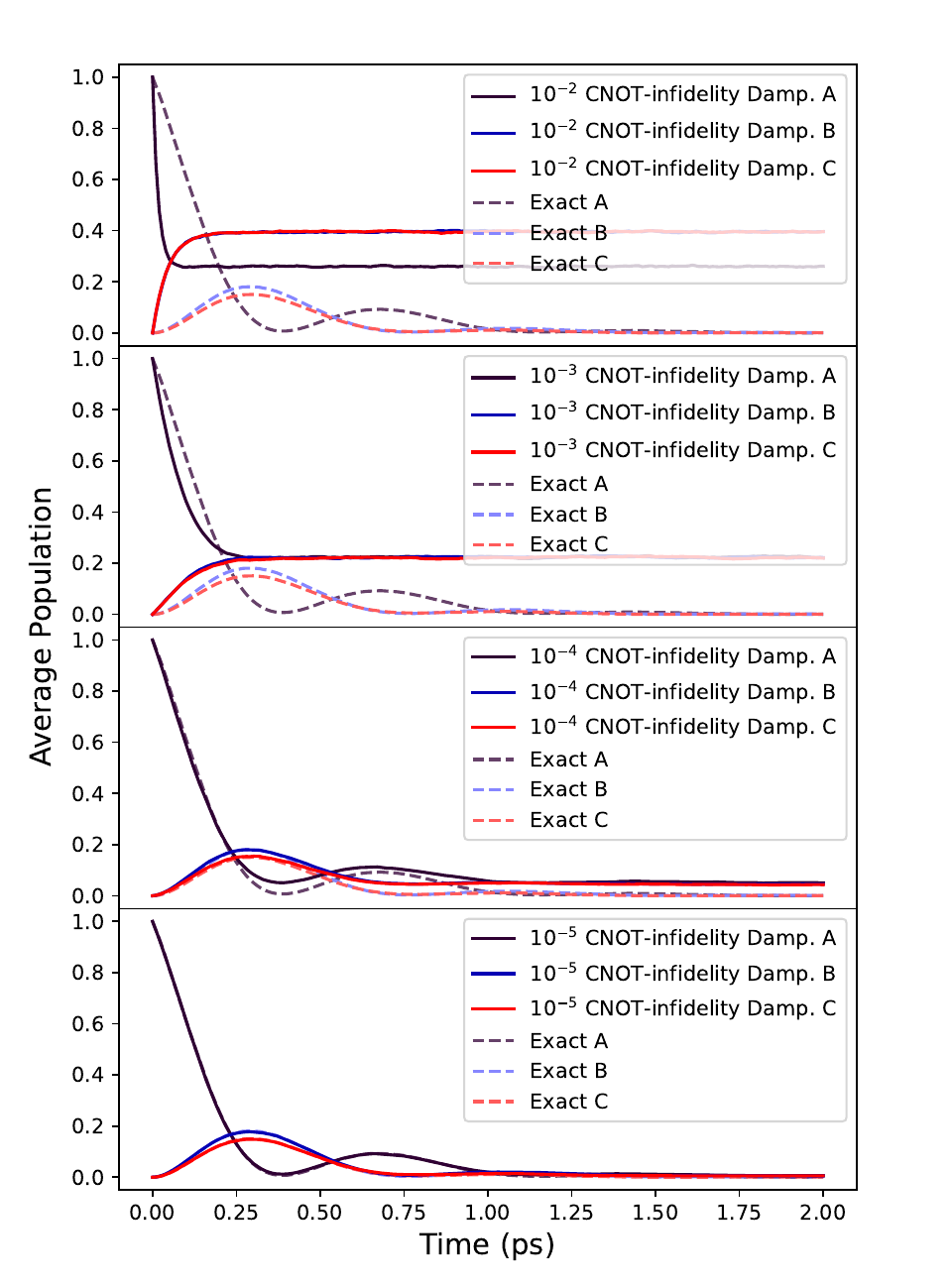}}
    \caption{(a) Population dynamics of the 3-site chromophore system under varying levels of CNOT infidelity, with $\varepsilon_{\rm CNOT} = 10^{-2}, 10^{-3}, 10^{-4}, 10^{-5}$. Each data point represents an average over 10,000 measurement shots. (b) Dissipative population dynamics of the 3-site chromophore system with amplitude damping under varying levels of CNOT infidelity. The error rates tested are $\varepsilon_{\rm CNOT} = 10^{-2}, 10^{-3}, 10^{-4}, 10^{-5}$. Each data point represents an average over 10,000 measurement shots.}
    \label{fig:3-site-CNOT-noisy}
\end{figure}

We conducted noise sweep simulations using {\scshape Qiskit}'s Noise Models module. Fig.~\ref{fig:3-site-CNOT-noisy} show the population dynamics of both pure and dissipative 3-site chromophore systems under various levels of CNOT infidelity.  This infidelity is modeled by an amplitude damping channel with error $\varepsilon_{\rm CNOT,amp}$ followed by a dephasing channel with error $\varepsilon_{\rm CNOT,dep}$. Based on the relative photon loss and dephasing rates in the qubit (Appendix~\ref{asec:fidelity-of-cD-gate}), we set $\varepsilon_{\rm CNOT} = \varepsilon_{\rm CNOT,amp} = 2\varepsilon_{\rm CNOT,dep}$ and analyze the noisy population dynamics for $\varepsilon_{\rm CNOT}$ values ranging from $10^{-2}$ to $10^{-5}$. Since each SWAP operation can be decomposed into three consecutive CNOT gates, the cumulative infidelity per SWAP operation is given by \begin{equation}
    \varepsilon_{\rm SWAP} \leq 1 - (1-\varepsilon_{\rm CNOT})^3.
\end{equation}

As expected, when $\varepsilon_{\rm CNOT}=10^{-5}$, the excited population dynamics closely match the ideal simulation. For larfer error rates, the effects of noise become more pronounced. Notably, at $\varepsilon_{\rm CNOT} = 10^{-4}$, the qualitative structure of population dynamics -particularly the relative excitation distribution among chromophores- remains discernible, albeit with some distortion. However, for $\varepsilon_{\rm CNOT} = 10^{-3}$, the noise overwhelms the system, rendering the dynamics unrecognizable. These results suggest that achieving a CNOT error rate of approximately $10^{-4}$ (0.01\% infidelity) or lower is essential for practical implementation of the chromophore dynamics simulation on circuit quantum electrodynamics (cQED) hardware.

\section{Estimating Fidelity, Idling Error of the Conditional Displacement Gate with Numerical Simulations}
\label{asec:fidelity-of-cD-gate}
The primary sources of infidelity in the conditional displacement (CD) gate arise from physical errors in both the cavity and qubit during the gate excution.  We model the composite system evolution under the Hamiltonian
{\allowdisplaybreaks\begin{equation}
    H_{\mathrm{CD}}/\hbar = \chi a^{\dag}a\frac{\sigma^{z}}{2} + \chi\left(\alpha a^{\dagger} + \alpha^{*}a\right)\sigma^{z},
    \label{eq:CD-ham}
\end{equation}}
where $\chi/2\pi \approx 50$ kHz is the weakly qubit-cavity dispersive coupling frequency, and $\alpha\leq 30$ is the displaced-frame amplitude to implement the CD operation~\cite{EickbuschECD} at a rate of $g_{\rm CD} = \alpha\chi$. 

{\em Error Sources}: We consider the following dominant sources of infidelity: photon loss in the cavity at a rate $\kappa_{1,c} \sim (1 \mathrm{ms})^{-1}$, photon loss in the qubit at a rate $\kappa_{1,q} \sim (100 \mu\mathrm{s})^{-1}$, qubit dephasing at a rate $\kappa_{\phi,q} \sim (200 \mu\mathrm{s})^{-1}$ (assuming the qubit has $T_{1} = T_{2}$ as a reasonable assumption), and qubit heating characterized by the thermal excited-state population $n_{\mathrm{th}} \approx 0.001 - 0.01$. Here $n_{\mathrm{th}}$ represents the steady-state of heating and loss. Together with $\kappa_{1,q}$, it fully describes the heating of the qubit and loss channels via detailed balance: $\kappa_{1,q} = \kappa_{\uparrow,q} + \kappa_{\downarrow,q}$, $(1-n_{th})\kappa_{\uparrow,q} = n_{th}\kappa_{\downarrow,q}$. These mechanisms apply to all idling times under the dispersive coupling Hamiltonian: 
\begin{align}
H/\hbar = \chi a^{\dag}a\frac{\sigma^{z}}{2}.    
\end{align}
In particular, qubit heating induces dephasing in the cavity at a characteristic rate of $\kappa_{\phi,c} \approx n_{th}\kappa_{1,q}$, which holds under the condition $\chi \gg \kappa_{1,q}$, ensuring that single loss or heating events fully dephase the cavity.~\cite{Reagor_memory_2016} We also note that the phase-flip ($\sigma^{z}$) errors on the qubit commute with the Hamiltonian. Therefore, they do not directly affect the fidelity of the CD gate itself but instead propagate to subsequent operations, i.e., we can either simulate the phase-flip error with the single-qubit $\sigma^z$ gate (assuming idle time) or perform quantum error correction.~\cite{Devitt2013QEC}

{\em Analysis of CD Gate Errors}: The probability of CD gate error, $\varepsilon_{\rm CD}$ can be estimated as: 
\begin{equation} \label{eq:asec_cd_error}
    \varepsilon_{\rm CD} = \kappa_{\rm all} \times \tau_{\rm gate}
\end{equation}
where $\kappa_{\rm all}$ is the total photon loss and dephasing rate (in Hz), and $\tau_{\rm gate}$ is the execution time of the CD gate on a physical quantum processor. 

To estimate $\kappa_{\rm all}$, we add up the rates in times per second for each of the four error sources mentioned above. As $\kappa_{\phi,c}$ is highly variable and at least 2 orders of magnitude less frequent than some of the other error rates, we can safely ignore cavity dephasing from our calculations for brevity to obtain $\kappa_{all}\approx 16~\text{kHz}$. The time necessary to perform the gate, $\tau_{gate}$, can also be calculated by dividing the displacement parameter by the CD operation rate ($g_{\rm CD} = \alpha\chi$) of the hardware. For each Trotter step, the displacement parameters are of the form $g_{cd,x}\tau/2$ (Appendix~\ref{asec:compiled_circuit_trotter}). Hence, Eq.~\eqref{eq:asec_cd_error} can be rewritten as 

\begin{align} \label{eq: CD_gate_error}
 \mathcal{E}_{CD} \approx (\kappa_{1,c} + \kappa_{1,q} + \kappa_{\phi,q} )\enspace\frac{g_{cd,x}  \tau}{2\alpha  \chi}.
\end{align}
\begin{figure}[htbp]
    \centering
    \subfloat[CD Gate Error]{\includegraphics[width=0.52\textwidth]{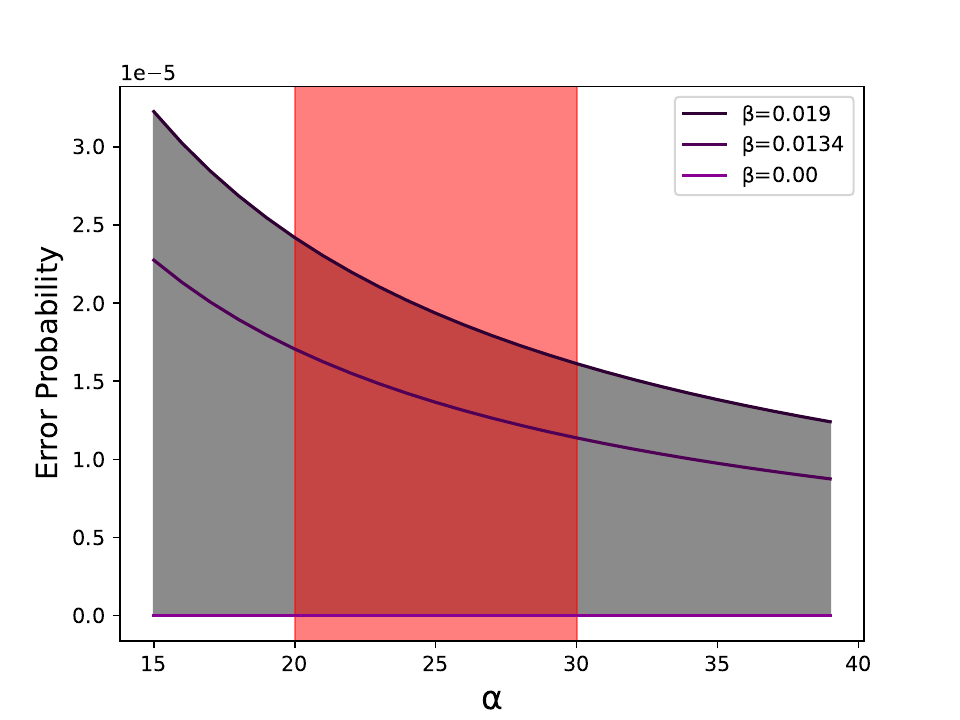}}
    \subfloat[CD Gate Noise]{\includegraphics[width=0.4\textwidth]{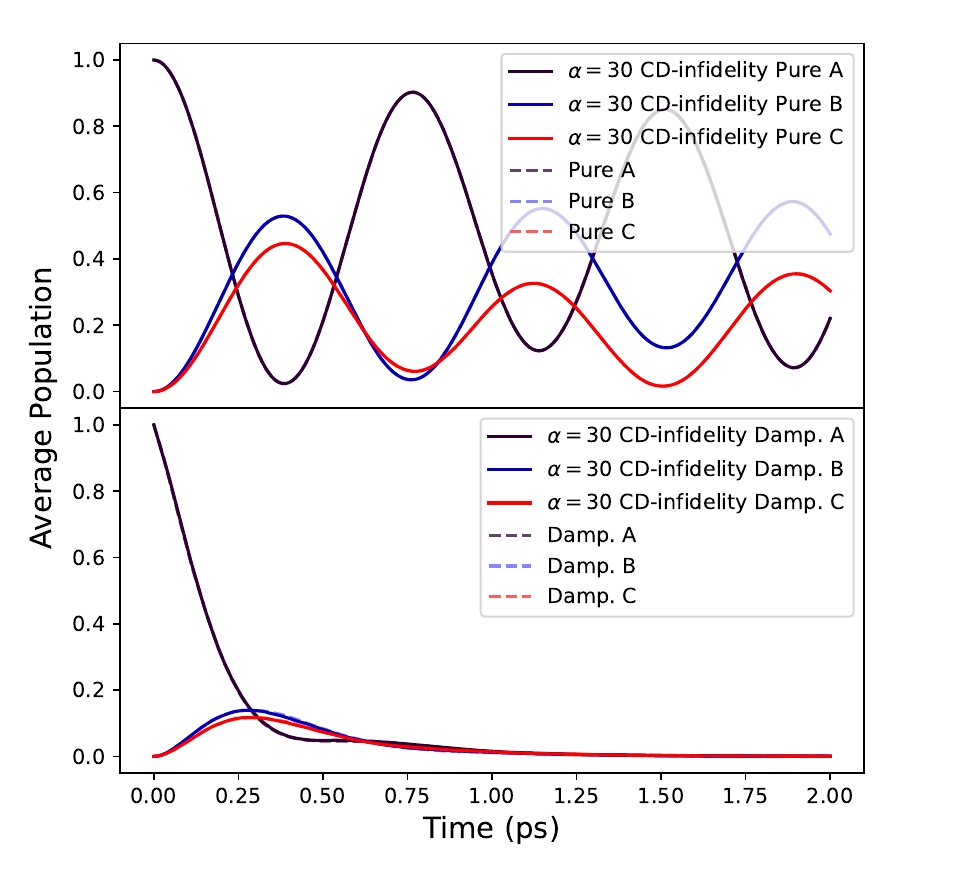}}
    \caption{(a) Calculated error probability for an individual CD($\beta$) gate with varying $\beta$ and a range of $\alpha = [15,40]$ defined in Eq.~\eqref{eq:CD-ham}. The red region captures realistic values of $\alpha$ that can be achieved on hardware, with an upper bound of $\alpha\leq 30$ (900 photons), whereas the gray region is indicative of the ranges of error probabilities that are possible for various values of $g_{cd,l}$. (b) Population dynamics of the 3-site chromophore system where the noisy CD gate's displaced-frame amplitude is $\alpha=30$. The top panel considers the non-dissipative system, whereas the bottom panel incorporates amplitude damping and dephasing with dissipative rates $\gamma_{\rm amp,all}$ and $\gamma_{\rm dep,all}$, respectively. 10,000 shots are performed for each case.}
    \label{fig:cd-err}
\end{figure}

For $x=l$, we present the range of expected error probabilities for our CD gate on chromophore $A$'s low-frequency cavity in Fig.~\ref{fig:cd-err} (a). These calculations are based on a Trotter step size $\tau= 10$ fs and various values of $S_l \in \{0.10, 0.05, 0.00\}$, which correspond to the CD rates $g_{cd,l} \in \{\text{1.90}\times 10^{12}, \text{1.34}\times 10^{12}, 0.00\}$, respectively. We observe that the error probabilities are relatively low, namely between $1.14\times 10^{-5}(\alpha = 30)$ and $1.71\times 10^{-5} (\alpha=20)$. However, it is important to keep in mind that these errors are per Trotter step and per chromophore, and thus can compound as we evolve the system further. While our error analysis only covers $\beta = g_{cd,l}$, other coupling strengths including $g_{cd,a}, g_{cd,b}, g_{cd,c}$ will also introduce additional error to the simulation results. We finally observe from Eq.~\eqref{eq: CD_gate_error} that probability error increases proportionally with the coupling strengths $g_{cd,x}$.

Finally, we perform vibronic simulations incorporating CD infidelity, as modeled using Eq.~\eqref{eq: CD_gate_error}, with $\alpha = 30$, which represents the maximum displaced-frame amplitude achievable on hardware. To account for this infidelity, Eq.~\eqref{eq: CD_gate_error} is compiled as one dephasing and two amplitude damping channels acting on the cavity and its auxiliary qubit with probabilities
{\allowdisplaybreaks\begin{align}
    p_{\rm CD,amp,q} = \kappa_{1,q} \frac{g_{cd,x}\tau}{2\alpha\chi}, \quad\quad p_{\rm CD,amp,c} = \kappa_{1,c} \frac{g_{cd,x}\tau}{2\alpha\chi}, \quad\quad p_{\rm CD,dep,q} = \kappa_{\phi,q} \frac{g_{cd,x}\tau}{2\alpha\chi}.
\end{align}}
For details on implementing amplitude damping and dephasing channels for qubits, we refer the readers to Appendix~\ref{asec:channel-derivation}, and for modeling Markovian amplitude damping in bosonic modes, Ref.~\citenum{BinomialCodes}.

The results shown in Fig.~\ref{fig:cd-err} (b) indicate minimal deviation between noisy and ideal simulations. This aligns with the analysis from Sec.~\ref{ssec:resource-est} where the number of CD gates per Trotter step is significantly smaller than that of CNOT operations, leading to negligible overall impact. Moreover, we observe that the cavity amplitude damping channels do not influence the population dynamics measured in the high-frequency qubits: the terms in Eqs.~\eqref{eq:final-ham-3site-h0}-\eqref{eq:final-ham-3site-h2yy}, when compiled into CD gates, only modify the phase of the controlled qubits.


\bibliography{ref}


\begin{tocentry}
\begin{center}

\includegraphics[width=0.9\columnwidth]{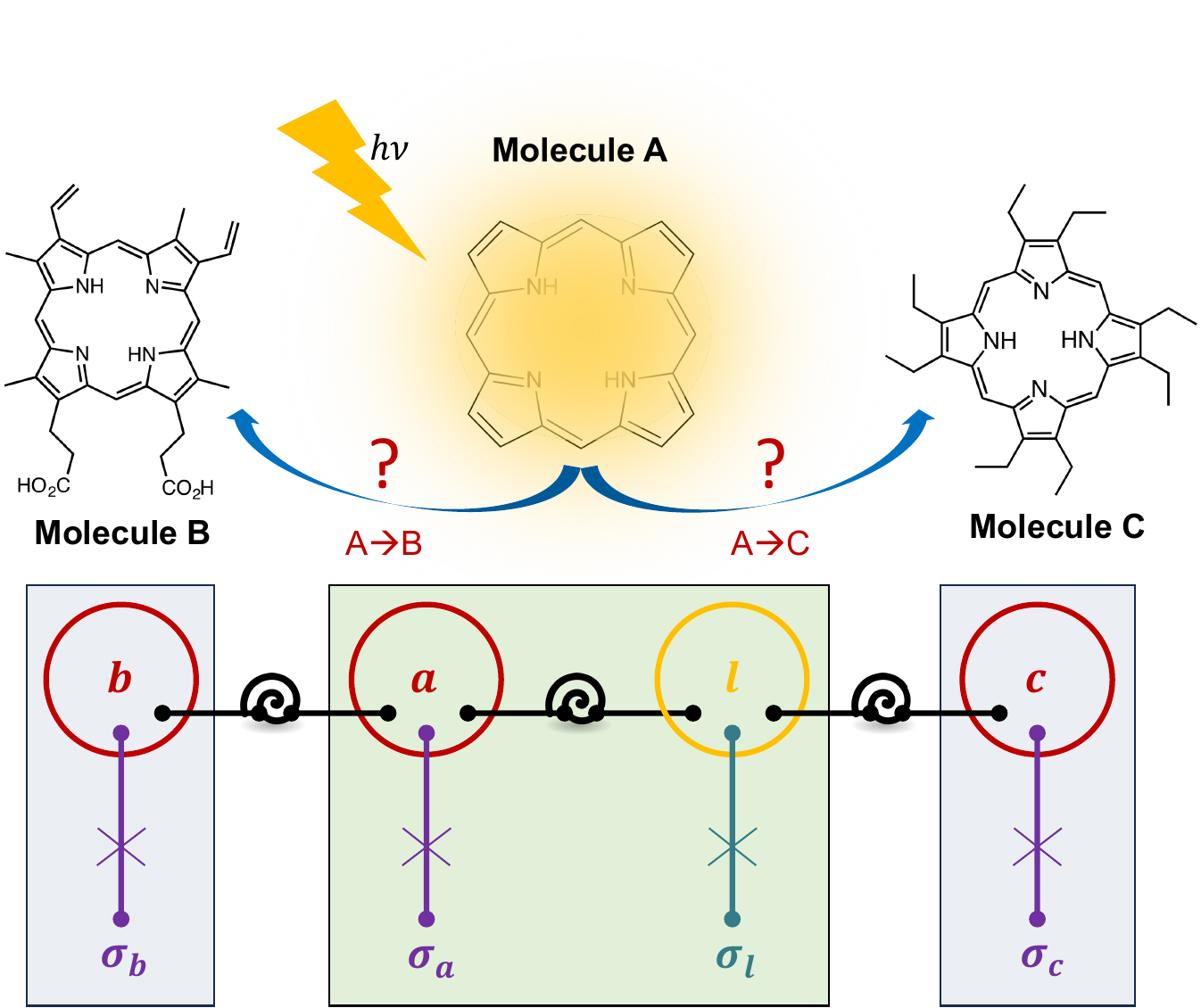}
    
\end{center}
\end{tocentry}

\end{document}